\documentclass[reprint,amsmath,amssymb,aps,prc,
superscriptaddress,
groupedaddress,
unsortedaddress,
runinaddress,
frontmatterverbose,
]{revtex4-1}
\usepackage{graphicx}
\usepackage{dcolumn}
\usepackage{bm}
\begin{document}
\preprint{arxiv: Sushant}
\title{Room temperature ferromagnetism in transparent and conducting\\ Mn-doped $SnO_{2}$ thin films}
\author{Sushant Gupta$^1$, V. Ganesan$^2$, N. P. Lalla$^2$, Indra Sulania$^3$ and B. Das$^1$}
\affiliation{$^1$Department of Physics, University of Lucknow, \\Lucknow-226007, India\\
$^2$UGC-DAE Consortium for Scientific Research, \\Indore-452017, India\\
$^3$Inter University Accelerator Centre (IUAC), \\New Delhi-110067, India}
\begin{abstract}
The magnetization as a function of magnetic field showed hysteretic behavior at room temperature. According to the temperature dependence of the magnetization, the Curie temperature ($T_{C}$) is higher than 350 K. Ferromagnetic Mn-doped tin oxide thin films exhibited low electrical resistivity  and high optical transmittance in the visible region (400-800 nm). The coexistence of ferromagnetism, high visible transparency and high electrical conductivity in the Mn-doped $SnO_{2}$ films is expected to be a desirable trait for spintronics devices.\\\\
\end{abstract}
\maketitle
\section{Introduction}
Tin oxide is an attractive material for solar cells and gas sensing applications due to its high optical transparency (above 80\% in the visible range of the electromagnetic spectra) and electrical conductivity (carrier concentration of the order of $10^{20}$ $cm^{-3}$) [1-4]. Recently there is an increased interest to introduce magnetic functionality in tin oxide semiconductors due to their promising applications in spintronics [5-14]. The tin oxide semiconductor can be made ferromagnetic by doping with transition-metal (TM) ions. The first report of high Curie temperature ferromagnetism in tin oxide thin films was by Ogale et. al. [15], who reported a Curie temperature $T_{c}$ = 650 K in pulsed laser deposited rutile $(Sn_{1-x}Co_{x})O_{2}$ thin films with x = 5-27\%, and an amazingly giant magnetic moment of (7.5$\pm$0.5)$\mu_{B}$ per Co ion. High Curie temperature ferromagnetism was latterly reported for $(Sn_{1-x}Ni_{x})O_{2}$ with x = 5\% [16, 17], $(Sn_{1-x}V_{x})O_{2}$ with x = 7\% [18], $(Sn_{1-x}Cr_{x})O_{2}$ with x = 5\% [19], and $(Sn_{1-x}Fe_{x})O_{2}$ with x = 14\% [20] \& x = 0.5-5\% [21]. Gopinadhan et. al. [22] investigated  $(Sn_{1-x}Mn_{x})O_{2}$ (with x = 10\%) thin films deposited by spray pyrolysis method and found ferromagnetic behavior above room temperature with low magnetic moment of 0.18$\pm$0.04 $\mu_{B}$ per Mn ion. Fitzgerald et. al. [23] studied 5\% Mn-doped $SnO_{2}$ bulk ceramic samples and reported a Curie temperature of $T_{c}$ = 340K with magnetic moment of 0.11 $\mu_{B}$ per ordered Mn ion. On the contrary, Duan et. al. [24]  reported an antiferromagnetic superexchange interaction in Mn-doped $SnO_{2}$ nanocrystalline powders and Kimura et. al. [25] observed paramagnetic behavior of Mn-doped $SnO_{2}$ thin films. Apart from this some other experiments were also carried out by various research groups on $SnO_{2}$ based dilute magnetic semiconductors (DMS) and reported interesting results regarding the absence or presence of ferromagnetism [26-50]. DMS based on $SnO_{2}$ could be useful for a variety of applications requiring combined optical and magnetic functionality [51]. Several devices such as spin transistors, spin light-emitting diodes, very high-density nonvolatile semiconductor memory, and optical emitters with polarized output have been proposed using $Sn_{1-x}(TM)_{x}O_{2}$ materials [52-55]. To investigate whether Mn is able to introduce thermodynamically stable high Curie temperature ferromagnetism in a $SnO_{2}$ semiconductor, $Sn_{1-x}Mn_{x}O_{2}$ (x = 0.000, 0.025, 0.050, 0.075, 0.100, 0.125, 0.150) thin films and  $Sn_{1-x}Mn_{x}O_{2}$ (x = 0.00, 0.01, 0.02, 0.03, 0.04, 0.05) bulk samples have been carefully prepared and characterized.
\section{Experimental details}
Thin films of $Sn_{1-x}Mn_{x}O_{2}$ with molar ratios of x = [Mn]/([Sn]+[Mn]) = 0.000, 0.025, 0.050, 0.075, 0.100, 0.125 and 0.150 were prepared by spray pyrolyzing a mixture of aqueous solutions of $SnCl_{4}.2H_{2}O$ and $(CH_{3}COO)_{2}Mn.4H_{2}O$ on glass substrates at $450^{o}C$.
An amount of 11.281 gm of $SnCl_{4}.2H_{2}O$ (Sigma Aldrich purity $>$ 99.99\%) was dissolved in 5 ml of concentrated hydrochloric acid by heating at $90^{o}C$ for 15 mins. The addition of hydrochloric acid rendered the solution transparent, mostly, due to the breakdown of the intermediate polymer molecules. The transparent solution subsequently diluted with ethyl alcohol formed the precursor. To achieve Mn doping, $(CH_{3}COO)_{2}Mn.4H_{2}O$ was dissolved in ethyl alcohol and added to the precursor solution. The amount of $(CH_{3}COO)_{2}Mn.4H_{2}O$ to be added depends on the desired doping concentration.
The doping concentration was varied from 0 to 15 at.\%. The overall amount of spray solution in each case was made together 50 ml. The repeated experiments of each deposition showed that the films could be reproduced easily. Pyrex glass slides (10 mm $\times$ 10 mm), cleaned with organic solvents, were used as substrates for various studies. During deposition, the solution flow rate was maintained at 0.2 ml/min by the nebulizer (particle size 0.5-10 $\mu$m). The distance between the spray nozzle and the substrate as well as the spray time was maintained at 3.0 cm and 15 min, respectively.

Bulk samples with nominal composition $Sn_{1-x}Mn_{x}O_{2}$ (where x = 0.00, 0.01, 0.02, 0.03, 0.04, 0.05) were synthesised by standard solid state reaction method. In the present investigation the bulk $SnO_{2}$ was synthesized by oxidizing the fine powder (50 mesh) of metallic tin (Sigma Aldrich purity $>$ 99.99\%) at 700$^{o}$C for 8 hrs in programmable temperature controlled SiC furnance. The appropriate ratio of the constituent oxides i.e. $SnO_{2}$ (as-synthesized) and $MnO_{2}$ (Sigma Aldrich purity $>$ 99.99\%) were throughly mixed and ground for several hours (4 to 6 hrs) with the help of mortar and pestle. After regrinding and mixing, the powder was kept in a alumina crucible and calcined at $750^{o}$C. After calcination the material was again ground to subdivide any aggregated products and to further enhance chemical homogeneity. These steps were repeated 3 to 4 times for better homogeneity and phase purity. The homogeneous powder thus formed was converetd into form of pellets before sintering. For this we employed the most widely used technique i.e. dry pressing, which consists of filling a die with powder and pressing at 400 kg/$cm^{2}$ into a compacted disc shape. In this way several cylindrical pellets of 2 mm thickness and 10 mm in diameter were prepared. Finally these pellets were put into alumina crucibles and sintered at about $1200^{o}$C in air for 16 hrs. The heating rate to the sintering temperature was about 100$^{o}$C/hour.

The gross structure and phase purity of thin films and bulk samples were examined by x-ray diffraction (XRD) technique using Bruker AXS (Model D8 Advanced, Germany) and Rigaku (Ultima IV, Japan) X-ray diffractometers. All the diffraction patterns were collected under a slow scan with a $0.01^{o}$ step size and a counting velocity of $0.5^{o}$ per minute. The experimental peak positions were compared with the data from the database Joint Committee on Powder Diffraction Standards (JCPDS) and Miller indices were assigned to these peaks. Hall measurements were conducted at room temperature to estimate the donor concentration (n), film resistivity ($\rho$) and carrier mobility ($\mu$) by using the van der Pauw geometry employing Keithley's Hall effect card and switching the main frame system. A specially designed Hall probe on a printed circuit board (PCB) was used to fix the sample to the size 10 $\times$ 10 $mm^{2}$. Silver paste was employed at the four contacts. The electrical resistivity and the sheet resistance of the samples were also determined using the four-point probe method with spring-loaded and equally spaced pins. The probe was connected to a Keithley voltmeter (2182A) \& constant-current source (2400) system and direct current and voltage were measured by slightly touching the tips of the probe on the surface of the samples. Multiple reading of current and the corresponding voltage were recorded in order to get average values. Atomic Force Microscopy (AFM) was performed with Multi Mode SPM (Digital Instrument Nanoscope E) in AFM mode to examine the microstructural evolution of the samples. Transmission Electron Microscopy (TEM) measurements were carried out on a Tecnai $20^{2}$G microscope with an accelerating voltage of 200 kV. All the images were digitally recorded with a slow scan charge-coupled device camera (image size 688 $\times$ 516 pixels), and image processing was carried out using the digital micrograph software. The TEM data were used for the study of grain size distribution and the crystalline character of the prepared samples. These TEM micrographs were also used to identify secondary phases present, if any, in the $Sn_{1-x}Mn_{x}O_{2}$ matrix. Optical absorption and transmission measurements were performed at room temperature within a wavelength range of 300-1100 nm using a Cary 5000 UV-Vis spectrophotometer having spectral resolution of 0.05 nm in the UV-Vis range. As a reference, 100\% baseline signals were displayed before each measurement. Magnetic measurements were carried out as a function of temperature (5 to 300 K) and magnetic field (0 to $\pm$2 T) using a `EverCool 7 Tesla' SQUID magnetometer. Measurements were carried out on small size samples placed in a clear plastic drinking straw. The data reported here were corrected for the background signal from the sample holder (clear plastic drinking straw) independent of magnetic field and temperature.
\section{Results and discussion}
\subsection{Structural properties}
\begin{figure}
  \includegraphics[height=11.5cm, width=8.3cm]{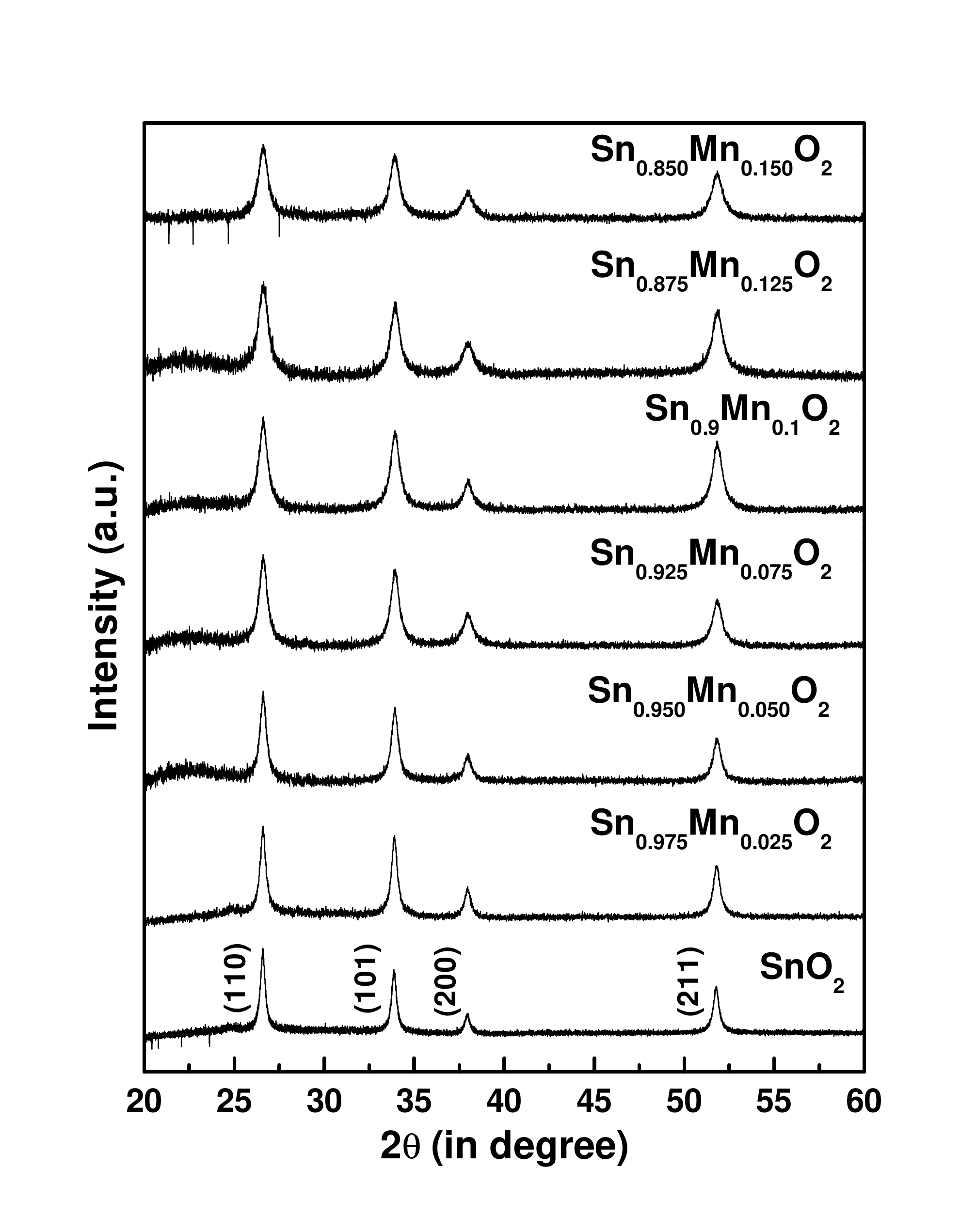}\\
  \caption{X-ray diffraction patterns of $Sn_{1-x}Mn_{x}O_{2}$ (x = 0.000, 0.025, 0.050, 0.075, 0.100, 0.125 and 0.150) thin films.}\label{3}
\end{figure}
XRD patterns of $Sn_{1-x}Mn_{x}O_{2}$ (x = 0.000, 0.025, 0.050, 0.075, 0.100, 0.125 and 0.150) thin films are shown in Fig. 1. It is evident that only the peaks corresponding to the rutile-type cassiterite phase of $SnO_{2}$ (space group $P4_{2}/mnm$) are detected with x up to 0.150. No additional reflection peaks related to impurities, such as unreacted manganese metal, oxides or any other tin manganese phases are detected. The lack of any impurity phases indicates that the Mn ion is incorporated well at the Sn lattice site. The lattice parameters (a and c) and cell volume ($a^{2}c$) were estimated using the (110), (101) and (200) peaks of $Sn_{1-x}Mn_{x}O_{2}$ for different doping concentration (x) and their average values are plotted in Fig. 2. The lattice parameters (a and c) and cell volume ($a^{2}c$) decrease with the increase in Mn doping concentration and reaches a minimum at doping level of 12.5 at.\% and for higher doping concentration ($\sim$ 15 at.\%) the same increases toward the value of pure $SnO_{2}$. An inflexion point is discernable between x = 0.050 and 0.075 (see Fig. 2), which could be attributed to the difference between the effective ionic radius of $Mn^{4+}$ (0.53 ${\AA}$, coordination number CN = 6) and high-spin $Mn^{3+}$ (0.645 ${\AA}$, CN = 6), while both are smaller than that of $Sn^{4+}$ (0.69 ${\AA}$, CN = 6) ions, i.e., Mn element acts as $Mn^{4+}$ upto x = 0.050 and as $Mn^{3+}$ in the x = 0.075, 0.100 and 0.125 films. The slope ratio of the line between x = 0.075 and 0.125 to that between x = 0 and 0.050 for cell volume is 25\%, which are comparable to the expected value of ($r_{Sn}^{4+}$ - $r_{Mn}^{3+}$)/($r_{Sn}^{4+}$ - $r_{Mn}^{4+}$) = 28\%, indicating the above interpretation is feasible. Above $\sim$ 12.5 at.\%, the observed lattice expansion indicates interstitial incorporation of Mn dopant ions. Interstitial incorporation of Mn ions might cause significant changes and disorder in the $SnO_{2}$ structure as well as many dramatic changes in the properties of the film, discussed in the following sections.

It is clear from Fig. 3 that Mn substitution effects the intensity of $SnO_{2}$ peaks. The normalized intensity i.e. $I_{110}(Sn_{1-x}Mn_{x}O_{2})/I_{110}(SnO_{2})$ decreases with doping, attains a minimum for 12.5\% doped sample, and afterwards increases. The intensity of the scattered x-ray is related to structure factor and this factor is determined by the presence of bound electrons in an atom. Since manganese has less bound electrons than tin, the substitution of tin by manganese in the tin oxide lattice should yield a structure factor, which is almost half the value for tin. If there is any change in the occupation site of manganese, i.e., substitutional ($Mn_{Sn}^{3+}$) to interstitial ($Mn_{i}$), the same may be reflected as an increase in the structure factor.
\begin{figure}
  \centering
  \includegraphics[height=7.1cm, width=8.7cm]{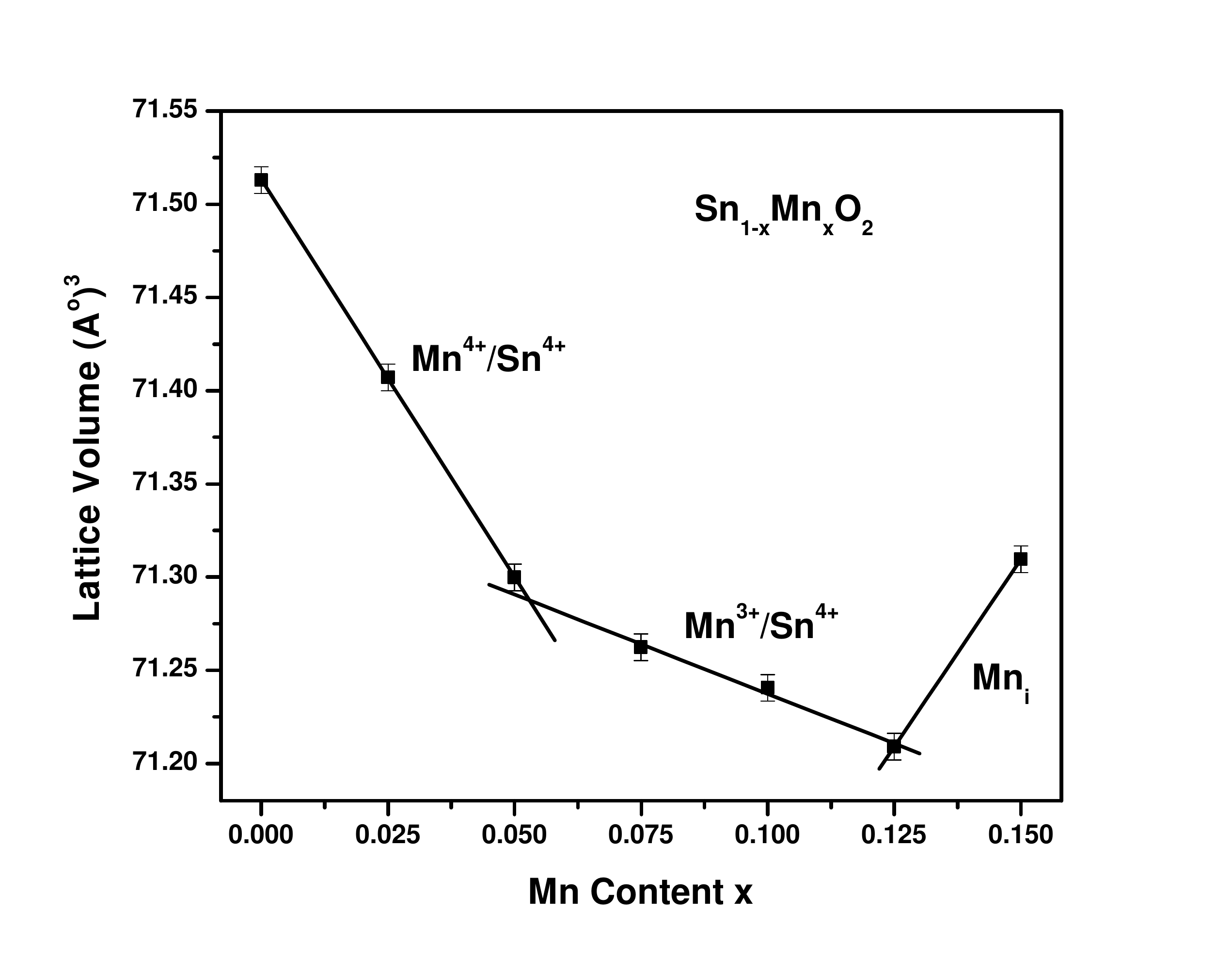}\\
  \caption{Variation of the lattice volume as a function of Mn concentration.}\label{4}
\end{figure}
\begin{figure}
  \centering
  \includegraphics[height=7.1cm, width=8.7cm]{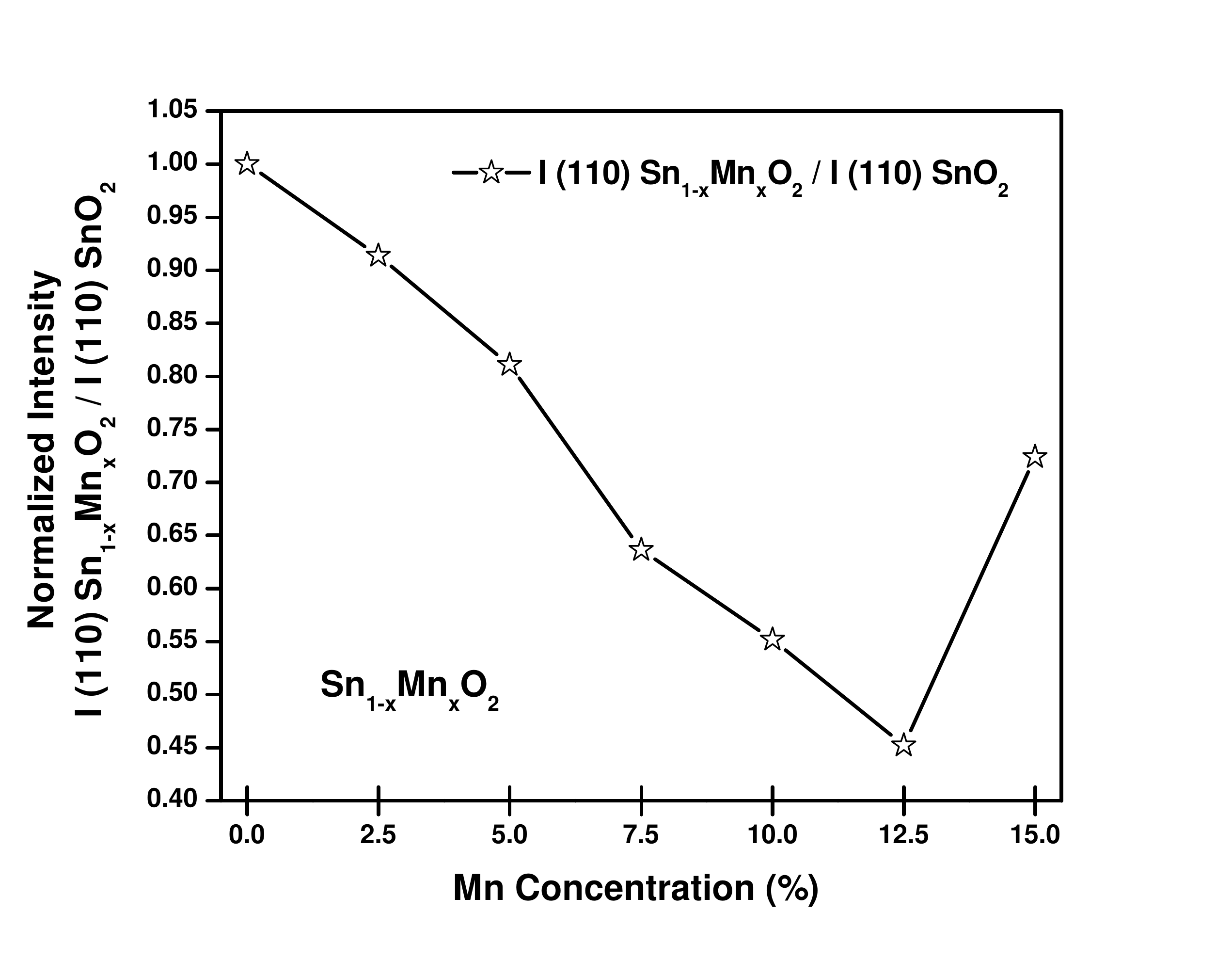}\\
  \caption{Changes in the (110) cassiterite peak intensity of respective thin films.}\label{4}
\end{figure}
\begin{table*}[htbp]
\centering
\renewcommand{\arraystretch}{1.6}
\caption{Lattice parameters, cell volume, texture coefficient C(hkl), the degree of preferential orientation $\sigma$, crystallite size and micro-strain for all the thin films.}
\vspace{3mm}
\begin{tabular}{||c|c|c|c|c|r|c|c|c|c|c||}
\hline
\hline
\centering
 & \multicolumn{2}{c|}{Lattice parameters}&Cell volume& \multicolumn{4}{c|}{Texture coefficient C(hkl)}& & Crystallite  & Non-Uniform  \\ \cline{2-3}\cline{5-8}
 Samples & a = b (${\AA}$)& c (${\AA}$)& (${\AA}^{3}$)& (110) & (101) & (200) & (211) & $\sigma$ & Size (nm) & Strain  \\ \hline \hline
  $SnO_{2}$& 4.7375& 3.1863& 71.5130& 0.976& 0.959& 1.081& 0.985& 0.048 & 39 & $0.935\times10^{-3}$ \\
\hline
$Sn_{0.975}Mn_{0.025}O_{2}$&4.7360& 3.1836& 71.4072& 0.831& 0.967& 1.339& 0.863& 0.202 & 30 & $1.009\times10^{-3}$ \\
\hline
$Sn_{0.950}Mn_{0.050}O_{2}$& 4.7340& 3.1815& 71.2998&0.902& 0.964& 1.337& 0.798& 0.203 & 26 & $1.304\times10^{-3}$  \\
\hline
$Sn_{0.925}Mn_{0.075}O_{2}$& 4.7335& 3.1805& 71.2624&0.829& 0.912& 1.472& 0.787& 0.276 & 19 & $1.159\times10^{-3}$ \\
\hline
$Sn_{0.9}Mn_{0.1}O_{2}$& 4.7333& 3.1798& 71.2406&0.786& 0.873& 1.271& 1.069& 0.187 & 17 & $0.859\times10^{-3}$ \\
\hline
$Sn_{0.875}Mn_{0.125}O_{2}$& 4.7327& 3.1792& 71.2091& 0.792& 0.809& 1.379& 1.021& 0.236 & 15 & $1.174\times10^{-3}$ \\
\hline
$Sn_{0.850}Mn_{0.150}O_{2}$& 4.7344& 3.1814& 71.3096&0.762& 0.870& 1.462& 0.906& 0.272 & 17 & $1.627\times10^{-3}$ \\
\hline
\hline
\end{tabular}
\end{table*}

In order to evaluate the effect of Mn doping on the average crystallite size and micro strain, the Williamson-Hall method [56, 57] was utilized by the equation
\begin{equation}\label{8}
  \beta \cos\theta = \frac{k\lambda}{D} + 4\varepsilon \sin\theta
\end{equation}
Where $\beta$ is the integral breadth of the peak from the (hkl) plane, k is a constant equal to 0.94, $\lambda$ is the wavelength of the radiation (1.5405 ${{\AA}}$ for $CuK_{\alpha}$ radiation), and $\theta$ is the peak position. The instrumental resolution in the scattering angle $2\theta$, $\beta_{inst}$, was determined by means of a standard crystalline silicon sample and approximated by
\begin{center}
$\beta_{inst}$ = $9\times10^{-6}$ $(2\theta)^{2}$ - 0.0005 $(2\theta)$ + 0.0623
\end{center}
Finally, the integral breadth $\beta$ without instrumental contribution was obtained according to the relation:
\begin{equation}\label{9}
  \beta  = \beta_{measured} - \beta_{instrumental}
\end{equation}
Eq. 1 represents the general form of a straight line $y = mx + c$. The plot between $\beta \cos\theta$ and $4 \sin\theta$ gives a straight line having slope $\varepsilon$ and intercept $k\lambda/D$. The values of crystallite size and micro-strain can be obtained from the inverse of intercept and the slope of the straight line, respectively. The Williamson-Hall plots for all the thin films are given in Fig. 4 and the results extracted from these plots are listed in Table I. The average particle size D decreases from $\sim$ 39 nm in pure $SnO_{2}$ to 26 nm in the sample with x=0.050. For higher doping concentration up to x=0.125, the particle size decreases further to $\sim$ 15 nm. This indicates that Mn incorporation hinders crystallite growth and the possible reason for this is the creation of Mn monolayer (Sn atoms are replaced by Mn atoms) on the surface of $SnO_{2}$ crystallite, which provides a barrier for surface diffussion and thus suppresses crystal growth [58-62]. In two-component materials, some of the constituents segregate to grain boundary, and lower the total Gibbs free energy [63].
\begin{figure}
  \centering
  \includegraphics[height=8.0cm, width=10.28cm]{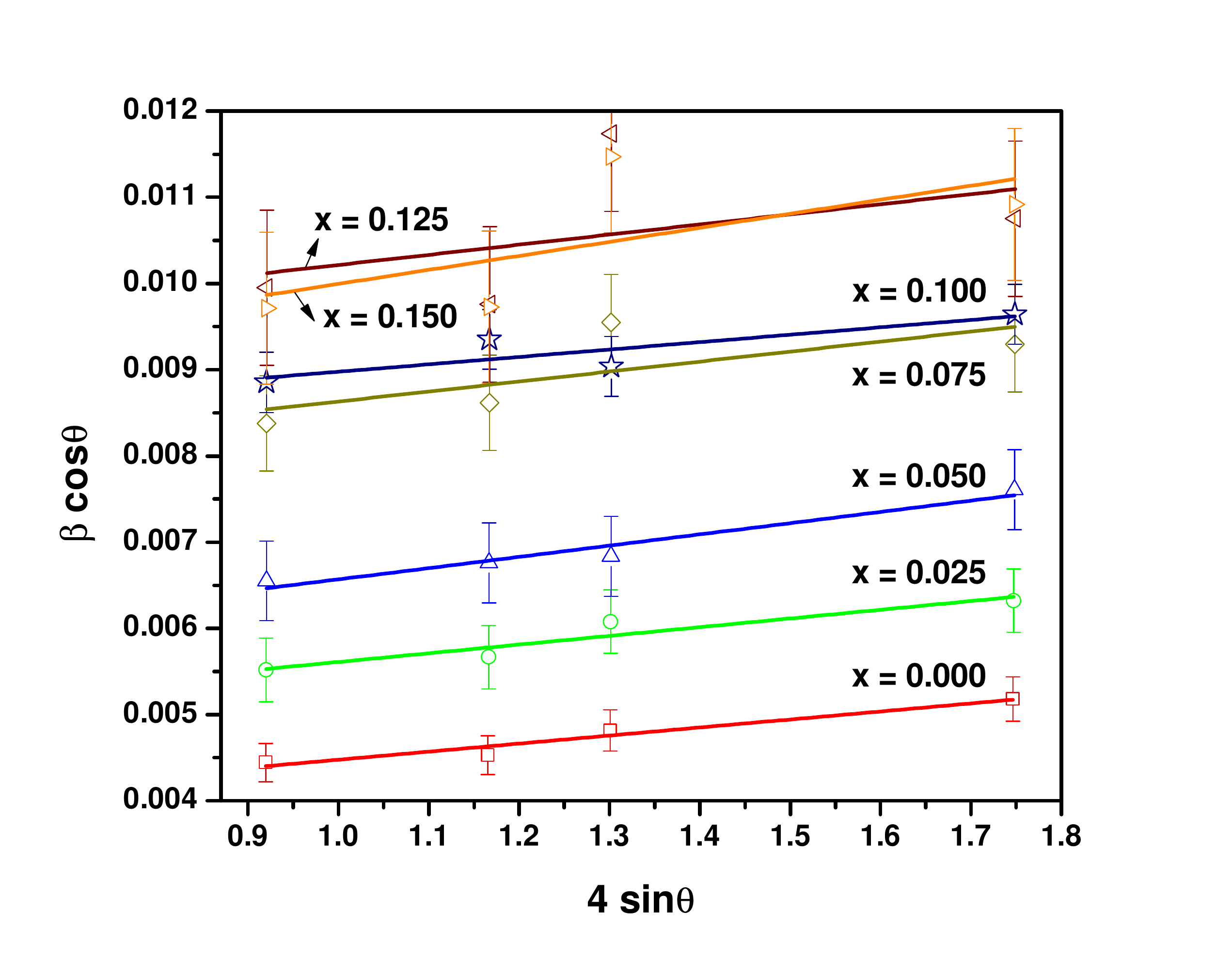}\\
  \caption{Williamson-Hall plots of all the thin films.}\label{6}
\end{figure}

The preferential orientation of the crystallites in the $Sn_{1-x}Mn_{x}O_{2}$ thin films was studied by calculating the texture coefficient C(hkl) of each XRD peak using the equation [64, 65]:
\begin{equation}\label{6}
  C(hkl) = \frac{N(I(hkl)/I_{o}(hkl))}{\sum (I(hkl)/I_{o}(hkl))}
\end{equation}
where $C (hkl)$ is the texture coefficient of the plane (hkl), I(hkl) is the measured integral intensity, $I_{o}(hkl)$ is the JCPDS standard integral intensity for the corresponding powder diffraction peak (hkl), and N is the number of reflections observed in the x-ray diffraction pattern. C(hkl) is unity for each XRD peak in the case of a randomly oriented film and values of C(hkl) greater than unity indicate preferred orientation of the crystallites in that particular direction. The degree of preferred orientation $\sigma$ of the film as a whole can be evaluated by estimating the standard deviation of all the calculated C(hkl) values [66]:
\begin{equation}\label{7}
  \sigma = \sqrt{\frac{\sum [C(hkl) - C_{o}(hkl)]^{2}}{N}}
\end{equation}
where $C_{o}(hkl)$ is the texture coefficient of the powder sample which is always unity.
The zero value of $\sigma$ indicates that the crystallites in the film are oriented randomly. The higher value of $\sigma$ indicates that the crystallites in the film are oriented preferentially [66]. The texture coefficient C(hkl) of all the XRD peaks along with the value of $\sigma$ for each $Sn_{1-x}Mn_{x}O_{2}$ film is given in Table I. It can be seen that the plane (200) has a high texture coefficient for all the films. The degree of preferred orientation $\sigma$ of the doped film is greater than that of pure $SnO_{2}$ film. However, it should be highlighted that none of the films possess a significant preferential orientation since the value of $\sigma$ is less than unity for all the films.
\begin{figure}
  \includegraphics[height=11.5cm, width=8.3cm]{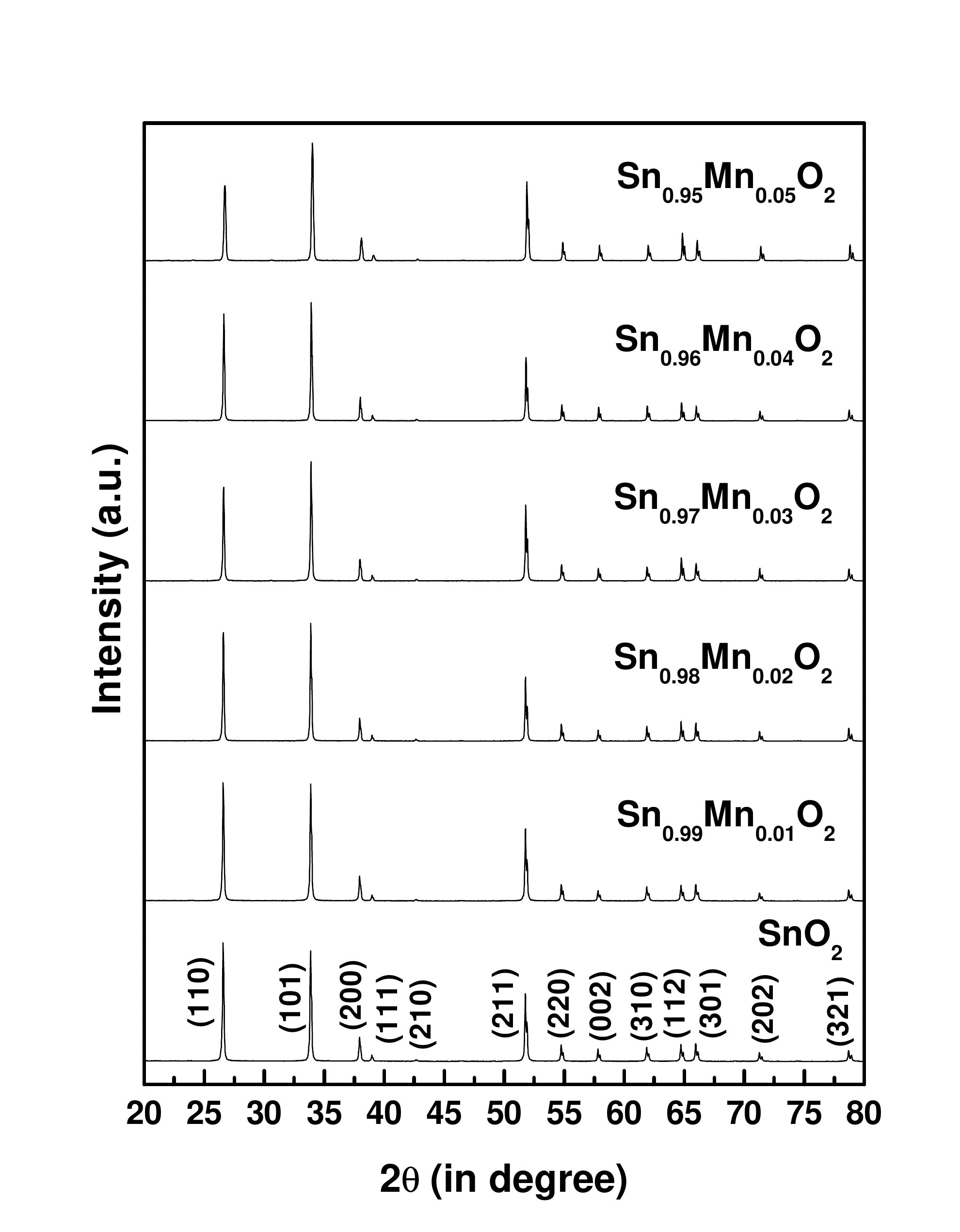}\\
  \caption{X-ray diffraction patterns of $Sn_{1-x}Mn_{x}O_{2}$ (x = 0.00, 0.01, 0.02, 0.03, 0.04 and 0.05) bulk samples.}\label{3}
\end{figure}

XRD patterns of $Sn_{1-x}Mn_{x}O_{2}$ (x = 0.00, 0.01, 0.02, 0.03, 0.04 and 0.05) bulk samples are shown in Fig. 5. The analysis of x-ray diffraction patterns reveals that all the bulk samples have a rutile-type cassiterite (tetragonal) phase of $SnO_{2}$, and the doping does not change the tetragonal structure (JCPDS \# 01-071-0652) of $SnO_{2}$. Furthermore, we could not find any diffraction peak corresponding to any impurity phase, such as unreacted Sn, Mn or other oxide phases, within the limit of instrumental sensitivity. We have calculated the lattice parameters using high angle XRD lines such as (301), (202) and (321) shown in Fig. 5. The determination of lattice constants of Mn doped $SnO_{2}$ bulk samples shows that on increasing the Mn concentration from 0 to 5 at.\%, the unit cell volume continuously reduces from its value for undoped $SnO_{2}$ samples as shown in Table II. The contraction of the lattice on Mn incorporation is thus a convincing evidence of the incorporation of Mn in $SnO_{2}$ lattice.  The effect of Mn doping on cell volume of bulk samples is opposite to that of thin films (Fig. 6). The slope ratio of the line between x = 0.00 and 0.03 to that between x = 0.04 and 0.05  for lattice volume is 23\%, which are comparable to the expected value of ($r_{Sn}^{4+}$ - $r_{Mn}^{3+}$)/($r_{Sn}^{4+}$ - $r_{Mn}^{4+}$) = 28\%, indicating Mn element acts as $Mn^{3+}$ upto x = 0.003 and as $Mn^{4+}$ in the x = 0.04 and 0.05 samples at room temperature.

Crystallite size of strain free $Sn_{1-x}Mn_{x}O_{2}$ bulk samples was calculated from x-ray diffraction data using the Debye-Scherrer formula:
\begin{equation}\label{9}
  D_{hkl} = \frac{0.9\lambda}{\beta \cos\theta}
\end{equation}
where $\lambda$ is the x-ray wavelength (1.5405 ${\AA}$ for $CuK_\alpha$), $\theta$ is the Bragg angle and $\beta$ is the full width of the diffraction line at half its maximum intensity (FWHM). The average crystallite sizes of $SnO_{2}$ doped with 0.0, 1.0, 2.0, 3.0, 4.0 and 5.0 at.\% of Mn which are calculated from Eq. (5) are 98, 93, 110, 107, 111 and 104 nm, respectively.
\begin{figure}
  \centering
  \includegraphics[height=7.1cm, width=8.7cm]{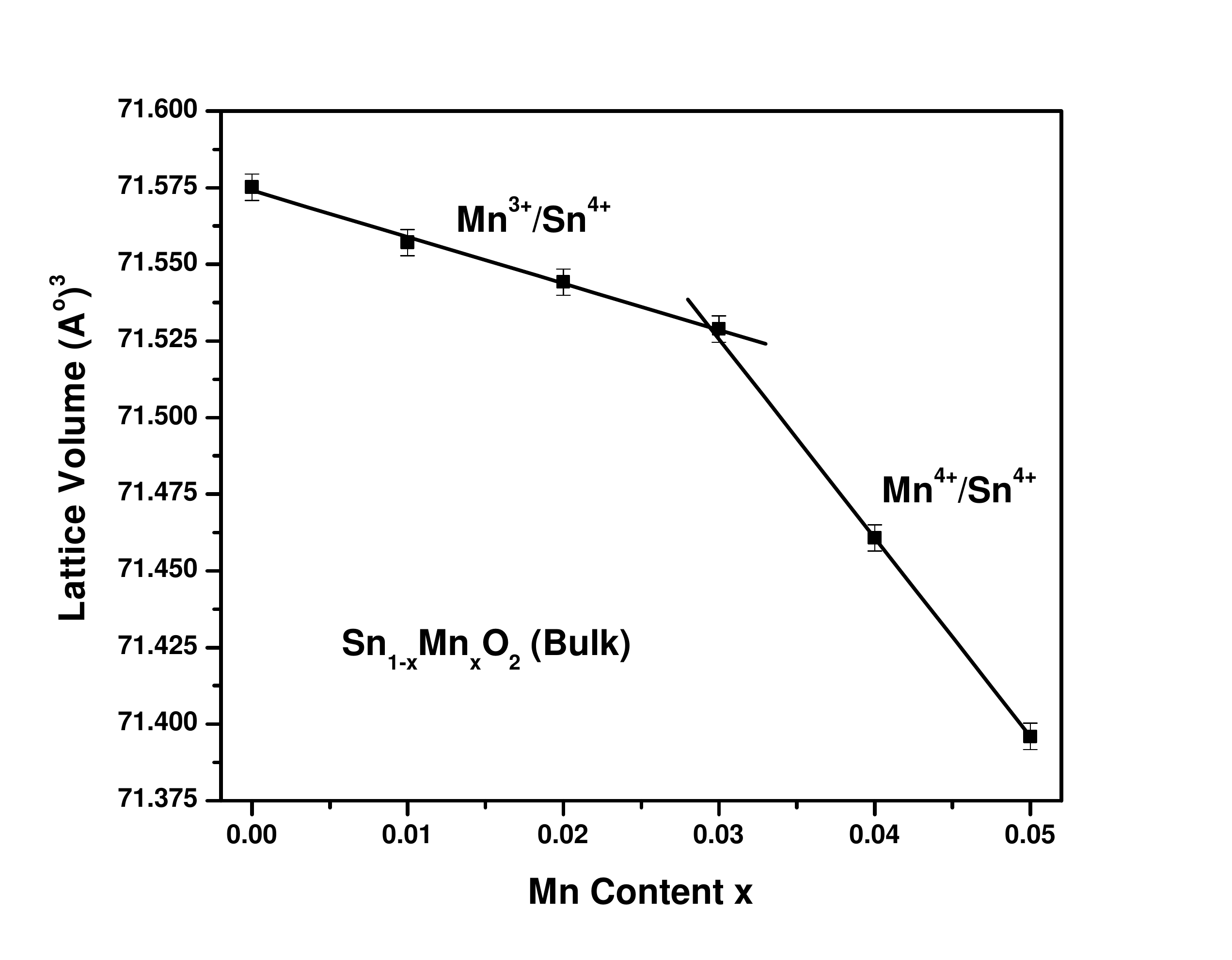}\\
  \caption{Variation of the lattice volume as a function of Mn concentration.}\label{4}
\end{figure}
\begin{table*}[htbp]
\centering
\renewcommand{\arraystretch}{1.6}
\caption{Lattice parameters, cell volume and crystallite size for all the bulk samples.}
\vspace{3mm}
\begin{tabular}{||c|c|c|c|c||}
\hline
\hline
\centering
 & \multicolumn{2}{c|}{Lattice parameters}&Cell volume&  Crystallite  \\ \cline{2-3}
 Samples & a = b (${\AA}$)& c (${\AA}$)& (${\AA}^{3}$)&  Size (nm) \\ \hline \hline
  $SnO_{2}$& 4.7377& 3.1888& 71.5752&  98 \\
\hline
$Sn_{0.99}Mn_{0.01}O_{2}$&4.7374& 3.1884& 71.5571& 93  \\
\hline
$Sn_{0.98}Mn_{0.02}O_{2}$& 4.7366& 3.1889& 71.5442& 110   \\
\hline
$Sn_{0.97}Mn_{0.03}O_{2}$& 4.7357& 3.1895& 71.5289& 107 \\
\hline
$Sn_{0.96}Mn_{0.04}O_{2}$& 4.7351& 3.1872& 71.4608& 111 \\
\hline
$Sn_{0.95}Mn_{0.05}O_{2}$& 4.7301& 3.1911& 71.3960& 104 \\
\hline
\hline
\end{tabular}
\end{table*}
\subsection{Microstructural properties}
In order to explore the effect of Mn doping on the microstructural characteristics in $Sn_{1-x}Mn_{x}O_{2}$ thin films, transmission electron microscopy was employed in imaging and diffraction modes. The transmission electron micrographs and the corresponding selected area electron diffraction (SAED) patterns for $Sn_{1-x}Mn_{x}O_{2}$ (x = 0.000, 0.025, 0.050, 0.075, 0.100 and 0.125) thin films are shown in Fig. 7. These TEM micrographs and SAED patterns have been analyzed using the IMAGE-J software. The TEM images of the $Sn_{1-x}Mn_{x}O_{2}$ (x = 0.000, 0.025, 0.050, 0.075, 0.100 and 0.125) films show the presence of interconnected nono-sized spheroidal grains. The crystallite size observed by TEM ($\sim$ 38 nm, 30 nm, 25 nm, 19 nm, 18 nm and 16 nm for x = 0.000, 0.025, 0.050, 0.075, 0.100 and 0.125, respectively) is in good agreement with that estimated by W-H plots ($\sim$ 39 nm, 30 nm, 26 nm, 19 nm, 17 nm and 15 nm for x = 0.000, 0.025, 0.050, 0.075, 0.100 and 0.125, respectively). The SAED patterns shown in Figs. 7(b), (d), (f), (h), (j) and (l) taken from $Sn_{1-x}Mn_{x}O_{2}$ (x = 0.000, 0.025, 0.050, 0.075, 0.100 and 0.125) films show several sharp rings, which are indexed to the (110), (101), (211) and (301) planes of the rutile crystalline structure of $SnO_{2}$. The electron diffraction pattern has been examined carefully for rings and spots of secondary phases, and it has been found that all the rings and spots belong to the tetragonal rutile structure of $SnO_{2}$ only. We have observed that there is no formation of any structural core-shell system.

AFM in tapping mode has been used to investigate the surface features of two bulk samples of different composition. The AFM images of $SnO_{2}$ and $Sn_{0.95}Mn_{0.05}O_{2}$ bulk samples are shown in Fig. 8. The size that is estimated from the few spherical grains is about 102 nm and 128 nm for x = 0.000 and 0.05, respectively. This is slightly larger than that obtained from XRD measurements.
\begin{figure*}
  \centering
  \includegraphics[height=10.6cm, width=18.5cm]{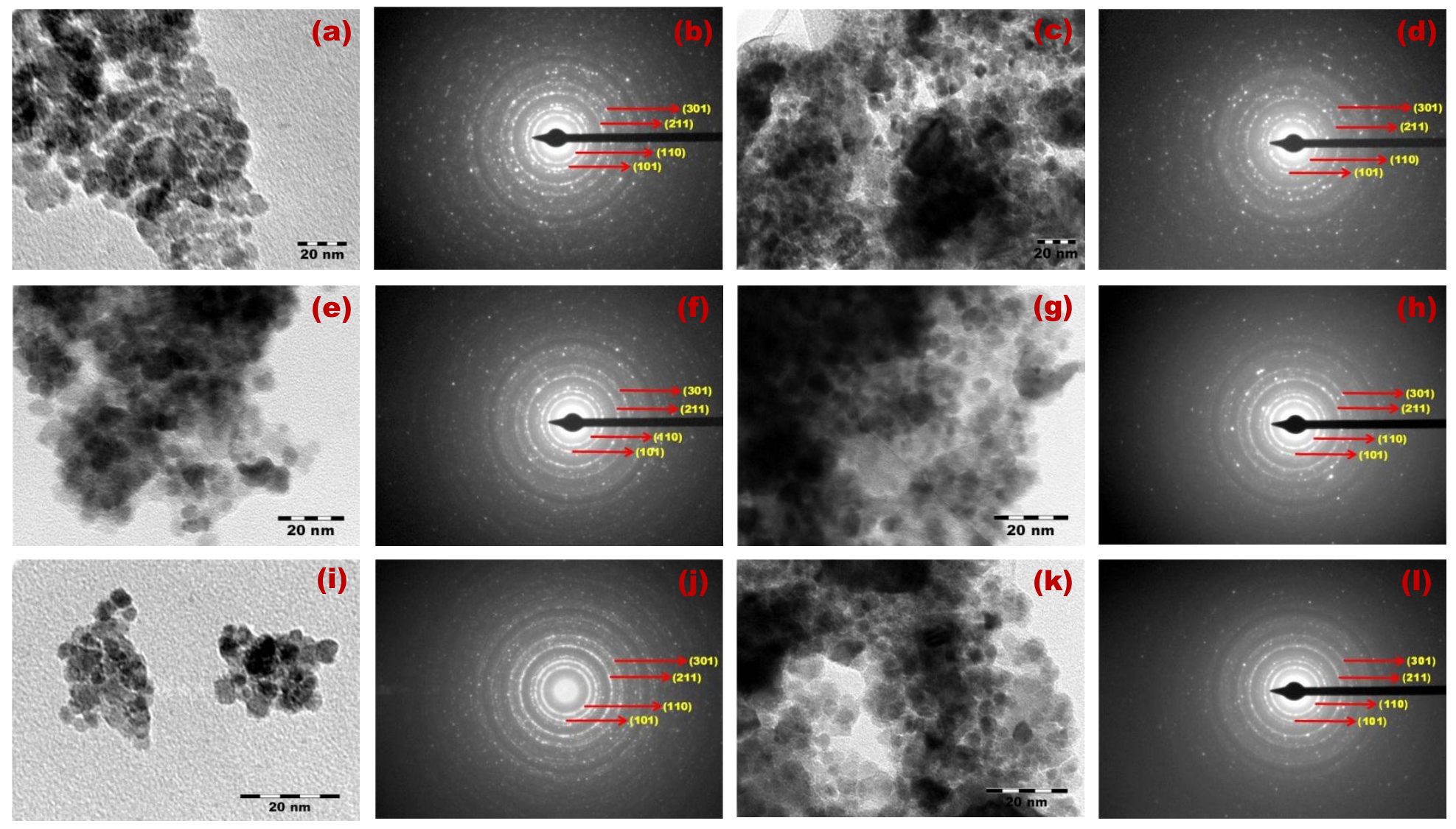}\\
  \caption{Transmission electron micrographs [(a), (c), (e), (g), (i) and (k)] of the $Sn_{1-x}Mn_{x}O_{2}$ (x = 0.000, 0.025, 0.050, 0.075, 0.100 and 0.125) thin films, respectively. Corresponding selected area electron diffraction (SAED) patterns for the $Sn_{1-x}Mn_{x}O_{2}$ (x = 0.000, 0.025, 0.050, 0.075, 0.100 and 0.125) thin films are shown in (b), (d), (f), (h), (j) and (l), respectively. Transmission electron micrographs of both thin films showing several nanocubes or nanospheres.}\label{4}
\end{figure*}
\begin{figure*}
  \centering
  \includegraphics[height=7.79cm, width=16cm]{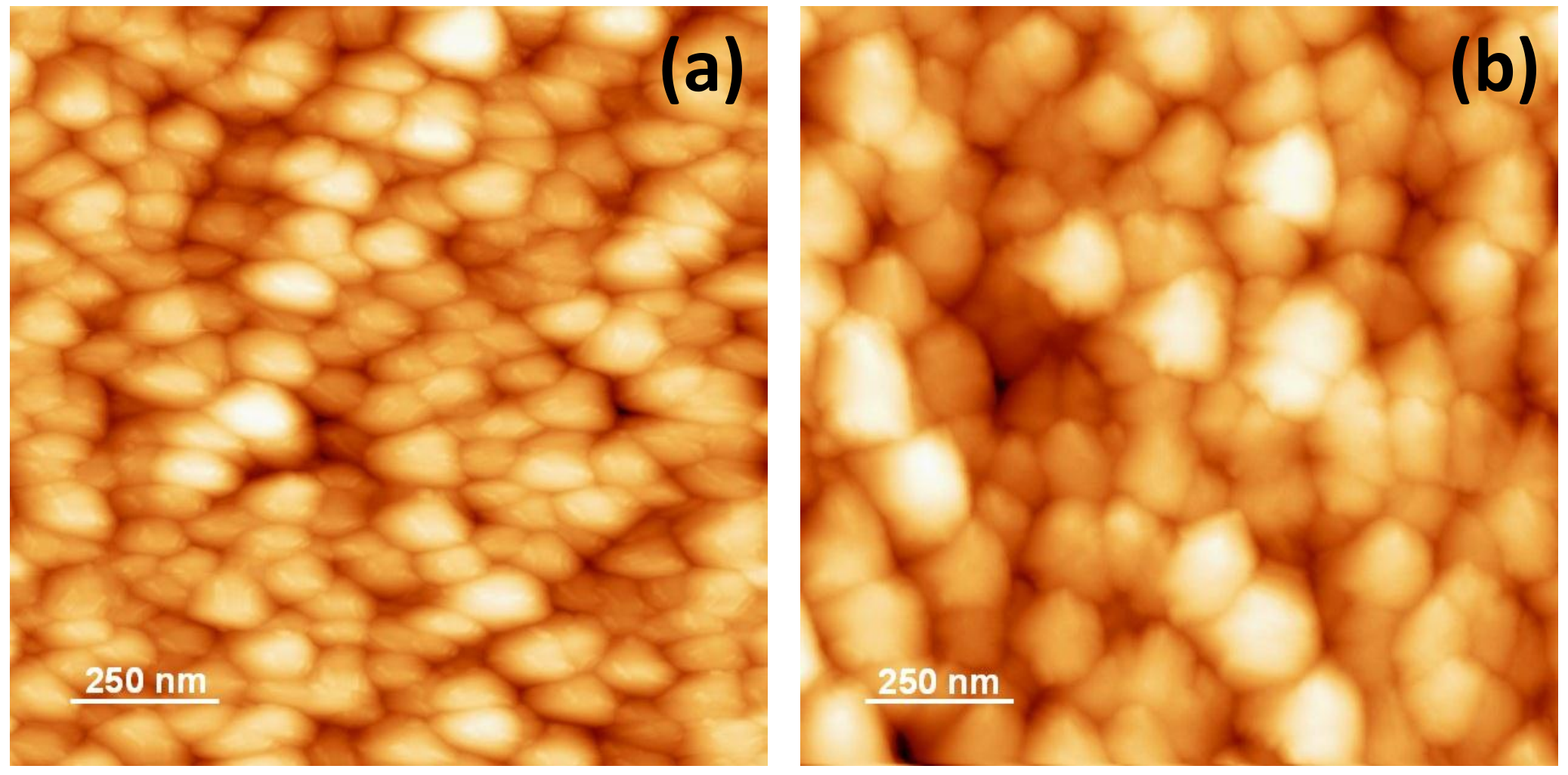}\\
  \caption{Atomic force microscopy images of (a) $SnO_{2}$ and (b) $Sn_{0.95}Mn_{0.05}O_{2}$ pellet surfaces.}\label{12}
\end{figure*}
\subsection{Electrical properties}
Transport properties (resistivity, carrier density, and mobility) of pure and Mn-doped $SnO_{2}$ films were measured by Hall effect measurements
using van der Pauw geometry. The room temperature results are presented for all measured films in Table III. The as-deposited $SnO_{2}$ films show the best combination of electrical properties as follows: resistivity ($\rho$) of 1.90 $\times$ $10^{-3}$ $\Omega$ cm, carrier concentration (n) of 2.704 $\times$ $10^{20}$ $cm^{-3}$, and mobility ($\mu$) of 12.165 $cm^{2}V^{-1}s^{-1}$. It has long been thought that native defects such as oxygen vacancy $V_{O}$ and tin interstitial $Sn_{i}$ are responsible for the observed n-type conductivity [67]. First principle calculations have provided evidence that usual suspects such as oxygen vacancy $V_{O}$ and tin interstitial $Sn_{i}$ are actually not responsible for n-type conductivity in majority of the cases [68-71]. These calculations indicate that the oxygen vacancies are a deep donor, whereas tin interstitials are too mobile to be stable at room temperature [68, 70]. Recent first principle calculations have drawn attention on the role of donor impurities in unintentional n-type conductivity [68-75]. Hydrogen is indeed a especially ambidextrous impurity in this respect, since it is extremely difficult to detect experimentally [68-71]. By means of density functional calculations it has been shown that hydrogen can substitute on an oxygen site and has a low formation energy and act as a shallow donor [68-71]. Hydrogen is by no means the only possible shallow donor impurity in tin oxide, but it is a very likely candidate for an impurity that can be unintentionally incorporated and can explain observed unintentional n-type conductivity [71]. Several groups have reported on the incorporation of hydrogen in tin oxide and many have claimed that hydrogen substitutes for oxygen [68-70, 76-90].
\begin{table*}[htbp]
\centering
\renewcommand{\arraystretch}{1.54}
\caption{Electrical parameters for all the thin films. }
\vspace{3mm}
\begin{tabular}{||c|c|c|c|c|c|c|c||}
\hline
\hline
\centering
 & Film & Sheet& & Carrier & Degeneracy& \multicolumn{2}{c||}{Carrier mobility}\\
 &thickness &resistance & Resistivity & concentration& temperature&\multicolumn{2}{c||}{$\mu$ ($cm^{2}V^{-1}s^{-1}$)}\\ \cline{7-8}
 Samples &  (nm)& $R_{s}$ ($\Omega/\square$) & $\rho$ ($\Omega$ cm) & n ($cm^{-3}$)& $T_{D}$ (K)&Observed & Calculated\\ \hline \hline
  $SnO_{2}$& 730& 26 & 1.90$\times10^{-3}$& $2.704\times10^{20}$& 5723 & 12.165 & 10.394\\
\hline
$Sn_{0.975}Mn_{0.025}O_{2}$& 740& 33 & 2.44$\times10^{-3}$& $2.251\times10^{20}$& 5065 & 11.379 & -\\
\hline
$Sn_{0.950}Mn_{0.050}O_{2}$& 710& 42 & 2.98$\times10^{-3}$& $2.135\times10^{20}$& 4889 & 9.823 & -\\
\hline
$Sn_{0.925}Mn_{0.075}O_{2}$& 750& 85 & 6.38$\times10^{-3}$& $1.353\times10^{20}$& 3607 & 7.240 & -\\
\hline
$Sn_{0.900}Mn_{0.100}O_{2}$& 805& 126 & 1.01$\times10^{-2}$& $1.149\times10^{20}$& 3235 & 5.386 & -\\
\hline
$Sn_{0.875}Mn_{0.125}O_{2}$& 790& 173 & 1.37$\times10^{-2}$& $1.225\times10^{20}$& 3376 & 3.724 & -\\
\hline
$Sn_{0.850}Mn_{0.150}O_{2}$& 776& 182 & 1.41$\times10^{-2}$& $1.032\times10^{20}$& 3011 & 4.295 & -\\
\hline
\hline
\end{tabular}
\end{table*}

The effect of Mn doping on the carrier concentration (n), resistivity ($\rho$) and Hall mobility ($\mu$) of the $SnO_{2}$ films are shown in Table III.
For the dopant concentration of 12.5 at.\% of Mn in $SnO_{2}$ the sheet resistance and resistivity is found maximum with the values of 1.73 $\times$ $10^{2}$ $\Omega$/sq. and 1.37 $\times$ $10^{-2}$ $\Omega$ cm, respectively.  It is apparent from the Table III that the carrier concentration and the mobility of the films show a consistent decrease with increase in Mn doping concentration (up to 12.5 at\%). Regarding electrical resistivity; although the undoped films show minimum resistivity, the resistivity of the doped film is still not too high compared to the reported values. The increase of the resistivity was mainly due to the decrease of the mobility. Beyond 5 at.\% of Mn  doping, another reason for increasing resistivity is attributed to the fact that $Mn^{3+}$ ions are substituted into $Sn^{4+}$ sites and act as an acceptor in $SnO_{2}$ lattice. This trend is accompanied by decrease of carrier concentration because of the presence of carrier traps.

The temperature dependence of electrical resistivity in the range 30-200$^{o}C$ indicates that the pure and Mn doped $SnO_{2}$ films are degenerate semiconductors. The film degeneracy was further confirmed by evaluating degeneracy temperature of the electron gas $T_{D}$ by the expression [91, 92]:
\begin{equation}\label{12}
  k_{B}T_{D} \simeq (\frac{\hbar^{2}}{2m^{*}})(3\pi^{2}n)^\frac{2}{3} = E_{F},
\end{equation}
where $m^{*}$ is the reduce effective mass and n is the electron concentration. The degeneracy temperature of all investigated films is clearly listed in  Table III. It can be seen that $T_{D}$ of $Sn_{1-x}Mn_{x}O_{2}$ (x = 0.000, 0.025, 0.050, 0.075, 0.100, 0.125 and 0.150) thin films are well above room temperature.

We have tried to identify the main scattering mechanisms that influence the mobility of pure $SnO_{2}$ films. There are many scattering mechanisms such as grain-boundary scattering, surface scattering, interface scattering, domain scattering, phonon scattering (lattice vibration), neutral, and ionized impurity scattering which may influence the mobility of the films [93, 94]. The interaction between the scattering centres and the carriers determines the actual value of the mobility of the carriers in the thin films.  In the interpretation of the mobility obtained for pure $SnO_{2}$ films, one has to deal with the problem of mixed scattering of carriers. To solve this problem, one has to identify the main scattering mechanism and then determine their contributions. The pure $SnO_{2}$ films prepared here are polycrystalline. They are composed of grains joined together by grain boundaries, which are transitional regions between different orientations of neighboring grains. These boundaries between grains play a significant role in the
scattering of charge carriers in polycrystalline thin films. The grain boundary scattering has an effect on the total mobility
only if the grain size is approximately of the same order as the mean free path of the charge carriers ($D \sim \lambda$). The mean free path
for the degenerate thin films can be calculated from known mobility ($\mu$) and carrier concentration (n) using the following expression [92, 94]:
\begin{equation}\label{13}
  \lambda = (3\pi^{2})^{\frac{1}{3}}(\frac{\hbar\mu}{e})n^{\frac{1}{3}},
\end{equation}
The mean free path value calculated for the pure $SnO_{2}$ film is 1.604 nm which is considerable shorter than crystallite size (D $\sim$ 38 nm) estimated using TEM micrograph. Moreover, the effect of crystallite interfaces is weaker in semiconductors, with n $\geq$ $10^{20}$ $cm^{-3}$, observed here, as a
consequence of the narrower depletion layer width at the interface between two grains [95]. Based on above discussion it is concluded that grain boundary scattering is not a dominant mechanism.

The mobility of the free carrier is not affected by surface scattering unless the mean free path is comparable to the film thickness [96]. Mean free path value calculated for the pure $SnO_{2}$ film is 1.604 nm, which is much smaller than the film thickness ($\sim$ 730 nm). Hence, surface scattering can be ruled out as the primary mechanism. Scattering by acoustical phonons [97] apparently plays a subordinate role in the pure $SnO_{2}$ films because no remarkable temperature dependence have been observed between 30 and $200^{o}$C. Moreover, neutral impurity scattering can be neglected because the neutral defect concentration is negligible in the pure $SnO_{2}$ films [92, 94]. Electron-electron scattering, as suggested to be important in Ref. 94, can also be neglected as it does not change the total electron momentum and thus not the mobility. In high crystalline $SnO_{2}$ films, scattering by dislocations and precipitation is expected to be of little importance [98].

Another scattering mechanism popular in unintentionally doped semiconductors is the ionized impurity scattering. According to the Brooks-Herring formula [99], the relaxation time for coupling to ionized impurities is in the degenerate case, given by
\begin{equation}\label{14}
  \tau_{i} = \frac{(2m^{*})^{\frac{1}{2}}(\epsilon_{o}\epsilon_{r})^{2}(E_{F})^\frac{3}{2}}{\pi e^{4}N_{i}f(x)},
\end{equation}
with $N_{i}$ the carrier concentration of ionized impurities and f(x) given by
\begin{equation}\label{15}
  f(x) = ln(1+x) - \frac{x}{1+x},
\end{equation}
with
\begin{equation}\label{16}
  x = \frac{8m^{*}E_{F}R_{S}^{2}}{\hbar^{2}},
\end{equation}
The screening radius $R_{S}$ is given by
\begin{equation}\label{17}
  R_{S} = (\frac{\hbar}{2e})(\frac{\epsilon_{o}\epsilon_{r}}{m^{*}})^{\frac{1}{2}}(\frac{\pi}{3N_{i}})^{\frac{1}{6}},
\end{equation}
where $\epsilon_{r}$ is the relative dielectric permittivity and $m^{*}$ is the effective mass of the carriers.
The mobility ($\mu$) is defined as
\begin{equation}\label{18}
\mu = \frac{e\tau}{m^{*}},
\end{equation}
Substitution of the $\tau_{i}$ expression [Eq. (8)] in Eq. (12) yields the expression for mobility due to ionized impurities as
\begin{equation}\label{19}
  \mu_{i} = \frac{(\frac{2}{m^{*}})^{\frac{1}{2}}(\epsilon_{o}\epsilon_{r})^{2}(E_{F})^{\frac{3}{2}}}{\pi e^{3}N_{i}f(x)},
\end{equation}
Since all the $H_{O}^{+}$ defects present in the pure $SnO_{2}$ films will be fully ionized at room temperature, impurity ion concentration will be equal to the free carrier concentration. Thus taking $N_{i}$ = n, $m^{*}$ = 0.31m, $\epsilon_{r}$ = 13.5 [100] and using Eq. (6) in Eq. (13) we get simplified form as
\begin{equation}\label{20}
  \mu _{i} = \frac{2.4232 \times 10^{-4}}{f(x)},
\end{equation}
with
\begin{equation}\label{21}
  x = 1.7942 \times 10^{-9} n^{\frac{1}{3}},
\end{equation}
The calculated mobility and measured mobility values for pure $SnO_{2}$ thin films are 10.394 and 12.165 $cm^{2}V^{-1}s^{-1}$, respectively, both are comparable to each other. This clearly indicates that the main scattering mechanism reducing the intra-grain mobility of the electrons in pure $SnO_{2}$ films is the ionized impurity scattering. Ionized impurity scattering with singly ionized $H_{O}^{+}$ donors  best describes the mobility of pure $SnO_{2}$ thin films. This finding supports our assumption that $H_{O}^{+}$ defect is source of conductivity in pure $SnO_{2}$ film.
\subsection{Optical properties}
The variation of the optical absorption coefficient ($\alpha$) with photon energy $h\nu$ was obtained using the absorbance data for various films. The absorption coefficient $\alpha$ may be written in terms of optical band gap $E_{g}$ and incident photon energy $h\nu$ as follows [101]:
\begin{equation}\label{29}
  \alpha = \frac{[A(h\nu - E_{g})^{n}]}{h\nu}
\end{equation}
where A is a constant which is different for different transitions indicated by different values of superscript n, and $E_{g}$ is the corresponding band gap. For direct transitions n = 1/2 or n = 2/3, while for indirect ones n = 2 or 3, depending on whether they are allowed or forbidden, respectively. The band gap can be deduced from a plot of $(\alpha h\nu)^{2}$ versus photon energy $(h\nu)$. Better linearity of these plots suggests that the films have direct band transition. The extrapolation of the linear portion of the $(\alpha h\nu)^{2}$ vs. $h\nu$ plot to $\alpha$ = 0 will give the band gap value of the films [102]. Fig. 9 shows such plots for all the thin films and the linear fits obtained for these plots are also depicted in the same figure.
\begin{figure}
  \centering
  \includegraphics[height=6.93cm, width=9cm]{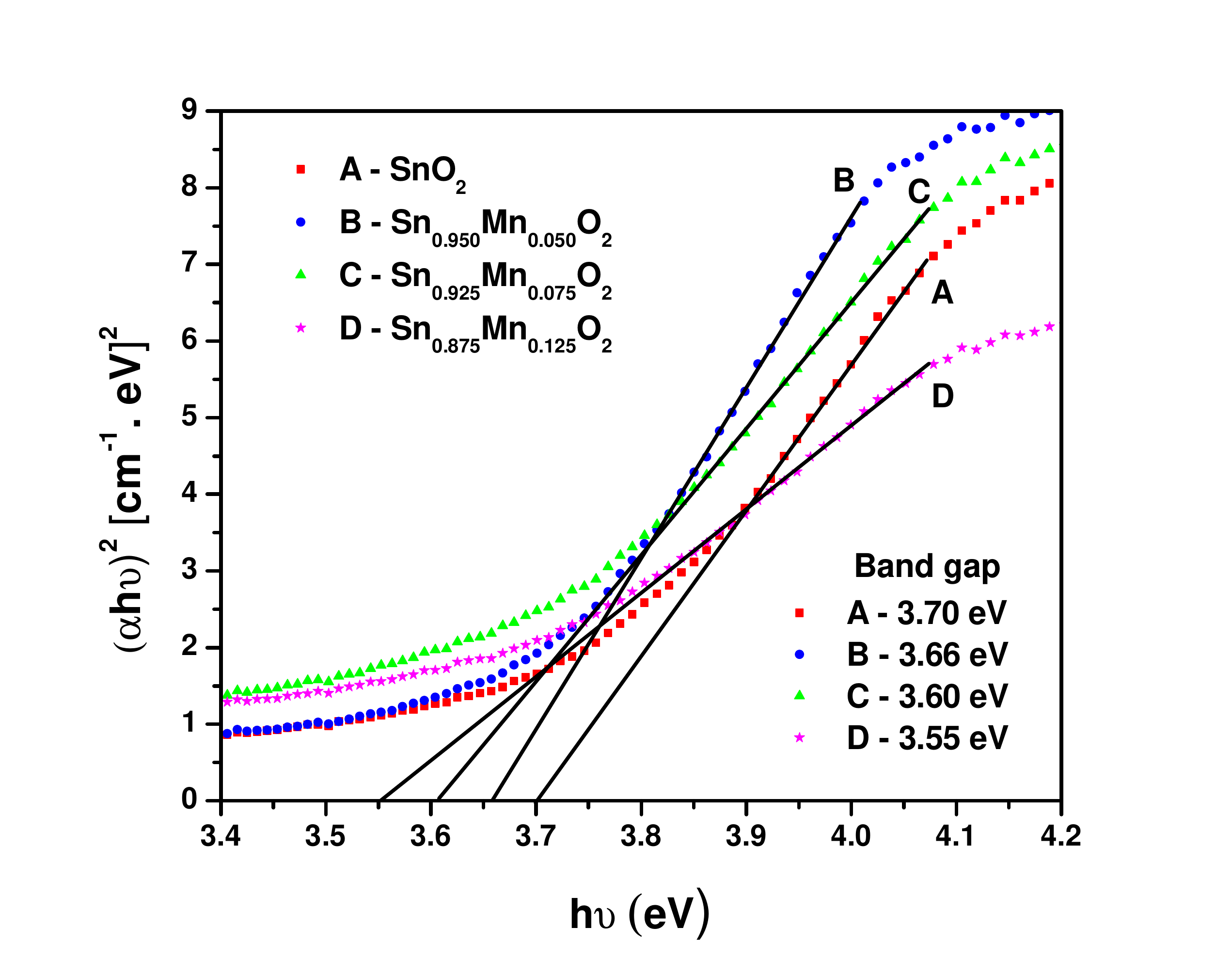}\\
  \caption{$(\alpha h\nu)^{2}$ vs $h\nu$ plots for the $Sn_{1-x}Mn_{x}O_{2}$ (x = 0.000, 0.050, 0.075 and 0.125) thin films. The direct energy band gap $E_{g}$ is obtained from the extrapolation to $\alpha$ = 0.}\label{14}
\end{figure}

With increasing Mn concentrations, the optical bandgap of the compounds shows a redshift compared to the host oxide $SnO_{2}$. We obtained the band gap to be 3.70 eV for pure $SnO_{2}$ and it starts decreasing for 5 at\%, 7.5 at\% and 12.5 at\% of Mn doped $SnO_{2}$ films as 3.66 eV, 3.60 eV and 3.55 eV, respectively. The decrease in bandgap for increasing Mn content is attributed to the strong exchange interactions between sp carriers of host $SnO_{2}$ and localized d electrons of Mn dopant [103].
\begin{figure}
  \centering
  \includegraphics[height=7.3cm, width=9.5cm]{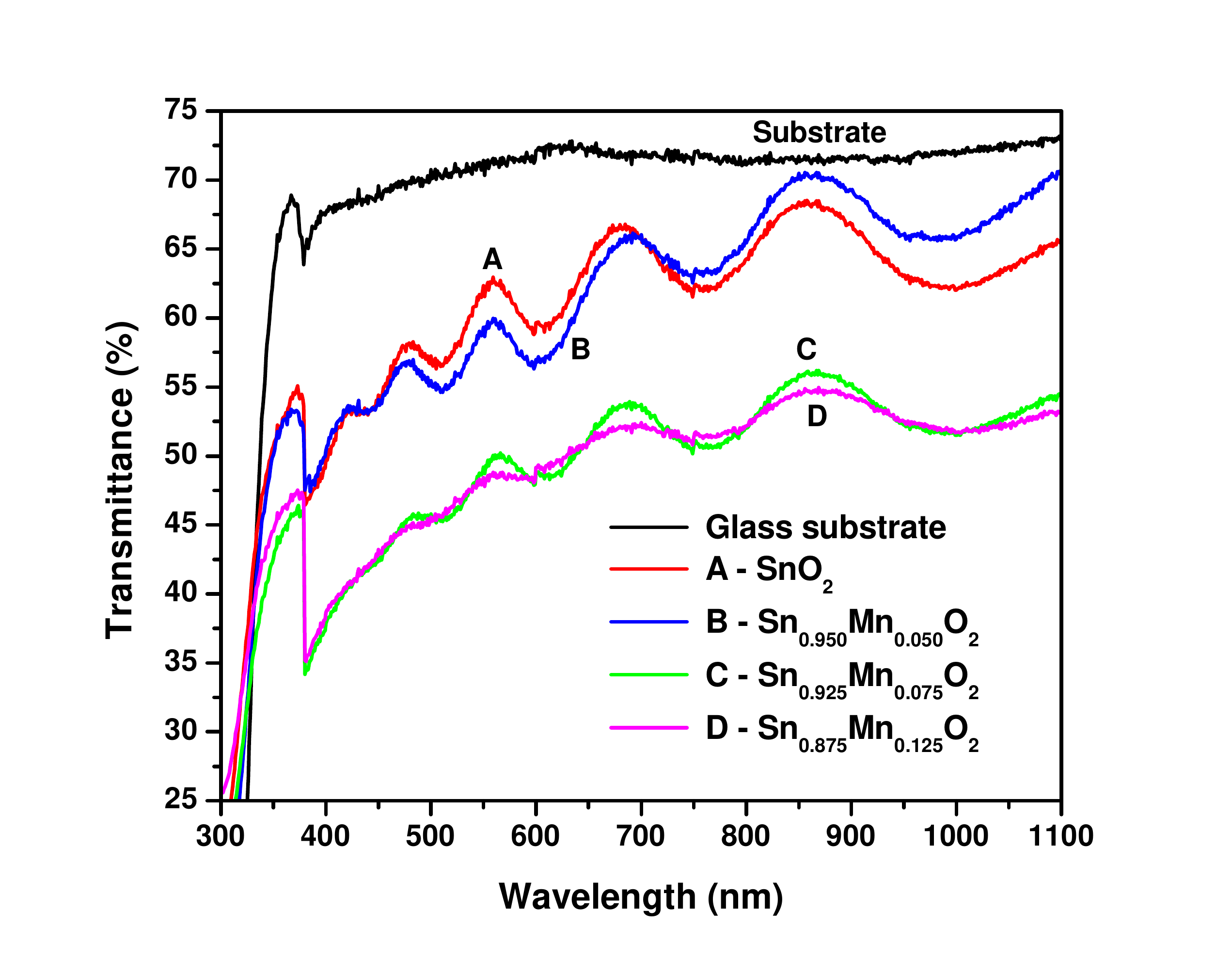}\\
  \caption{Comparison of transmittance spectra obtained from the different thin films.}\label{16}
\end{figure}

The optical transmittance spectra of uncoated glass substrate and Mn-$SnO_{2}$ ([Mn]/[Sn] = 0.0, 5.0, 7.5, 12.5 at.\%) thin films as a function of
wavelength ranging from 300 to 1100 nm is shown in Fig. 10. The transmittance exhibits interference in the visible range. Normally, fringes appear in the transmittance spectra due to the interference of the light, reflected between air-film and film-substrate interfaces. But, if the thickness is not uniform or slightly tapered then interference fringes may disappear from the transmission curve. The average transmittance of the Mn-doped $SnO_{2}$ films ranges from 50\% to 65\%. The thickness of as deposited films is approximately 700 nm, which is higher compared to the reported film thickness; this higher thickness affects the optical transmission. In the visible region, the average transmittance for the pure $SnO_{2}$ film is of $\sim$ 60\%, which is decreased upto $\sim$ 48\%  on addition of 12.5 at.\% of Mn. This relatively large absorption in $Sn_{1-x}Mn_{x}O_{2}$ (x = 0.075, 0.125) films may be the result of mixed valence of Mn ($Mn^{4+}/Mn^{3+}$). The presence of the $Mn^{3+}$ and $Mn^{4+}$ produces colour centres due to unpaired electrons in the d orbital that causes intense and deep coloration. Furthermore, the sharp decrease in transmittance at the shorter wavelength is as a result of the inter-band transition.
\begin{table*}[htbp]
\centering
\renewcommand{\arraystretch}{1.8}
\caption{Values of $\lambda$, $T_{M}$ and $T_{m}$ for $Sn_{1-x}Mn_{x}O_{2}$ (x = 0.000, 0.050, 0.075, 0.125) thin films corresponding to transmission spectra. The calculated values of refractive index and film thickness are based on the envelope method.}
\vspace{3mm}
\centering
\begin{tabular}{||c|c|c|c|c|c|c|c|c|c||c|c|c|c|c|c|c|c|c|c||} \hline\hline
\multicolumn{10}{||c||}{\bf{$SnO_{2}$}} &
\multicolumn{10}{c||}{\textbf{$Sn_{0.950}Mn_{0.050}O_{2}$}}\\ \hline
$\lambda$& $T_{M}$& $T_{m}$& $T_{s}$& s& n& $d_{pre}$& $m_{est}$& $m_{exact}$& $d_{accu}$& $\lambda$& $T_{M}$& $T_{m}$& $T_{s}$& s& n& $d_{pre}$& $m_{est}$& $m_{exact}$& $d_{accu}$\\
(nm)& & & & & & (nm)& & & (nm)& (nm)& & & & & & (nm)& & & (nm)\\ \hline
481& 0.578& 0.554& 0.697& 2.094& 2.272& -& 6.859& 7.0& 741& 478& 0.568& 0.541& 0.700& 2.083& 2.290& -& 6.688& 7.0& 731\\ \hline
506& 0.600& 0.569& 0.707& 2.058& 2.272& 793& 6.520& 6.5& 724& 509& 0.580& 0.546& 0.702& 2.076& 2.325& 862& 6.377& 6.5& 712\\ \hline
561& 0.626& 0.584& 0.715& 2.030& 2.296& 743& 5.943& 6.0& 733& 560& 0.600& 0.554& 0.711& 2.044& 2.358& 792& 5.878& 6.0& 712\\ \hline
603& 0.647& 0.596& 0.724& 1.999& 2.302& 714& 5.543& 5.5& 720& 600& 0.619& 0.564& 0.721& 2.009& 2.362& 607& 5.496& 5.5& 699\\ \hline
683& 0.665& 0.612& 0.718& 2.019& 2.317& 681& 4.926& 5.0& 737& 694& 0.662& 0.602& 0.720& 2.012& 2.351& 573& 4.729& 5.0& 738\\ \hline
757& 0.678& 0.621& 0.716& 2.026& 2.334& 742& 4.477& 4.5& 730& 753& 0.682& 0.627& 0.720& 2.012& 2.307& 680& 4.277& 4.5& 734\\ \hline
861& 0.682& 0.625& 0.713& 2.037& 2.341& 683& 3.948& 4.0& 736& 866& 0.705& 0.652& 0.715& 2.030& 2.297& 676& 3.703& 4.0& 754\\ \hline
994& 0.683& 0.622& 0.720& 2.012& 2.337& -& 3.414& 3.5& 744& 978& 0.708& 0.656& 0.720& 2.012& 2.273& -& 3.244& 3.5& 753\\ \hline
\multicolumn{10}{||c||}{$d_{pre}$(avg) = 726 nm, $d_{accu}$(avg) = 733 nm, d(exp) = 730 nm} &
\multicolumn{10}{c||}{$d_{1}$(avg) = 698 nm, $d_{2}$(avg) = 729 nm, d(exp) = 710 nm}\\ \hline\hline
\multicolumn{10}{||c||}{\textbf{$Sn_{0.925}Mn_{0.075}O_{2}$}} &
\multicolumn{10}{c||}{\textbf{$Sn_{0.875}Mn_{0.125}O_{2}$}}\\ \hline
$\lambda$& $T_{M}$& $T_{m}$& $T_{s}$& s& n& $d_{pre}$& $m_{est}$& $m_{exact}$& $d_{accu}$& $\lambda$& $T_{M}$& $T_{m}$& $T_{s}$& s& n& $d_{pre}$& $m_{est}$& $m_{exact}$& $d_{accu}$\\
(nm)& & & & & & (nm)& & & (nm)& (nm)& & & & & & (nm)& & & (nm)\\ \hline
510& 0.472& 0.454& 0.702& 2.076& 2.275& - & 6.620& 6.5& 729& 569& 0.487& 0.480& 0.715& 2.030 &2.106& -& 6.048& 6.0& 811\\ \hline
566& 0.502& 0.470& 0.716& 2.026& 2.335& 780& 6.122& 6.0& 727& 581& 0.492& 0.484& 0.715& 2.030 &2.115& 849& 5.948& 5.5& 755\\ \hline
604& 0.518& 0.485& 0.722& 2.006& 2.307& 755& 5.668& 5.5& 720& 695& 0.520& 0.505& 0.717& 2.023& 2.163& 654& 5.085& 5.0& 803\\ \hline
689& 0.540& 0.495& 0.720& 2.012& 2.386& 707& 5.139& 5.0& 722& 752& 0.534& 0.517& 0.720& 2.012& 2.163& 981& 4.700& 4.5& 782\\ \hline
760& 0.551& 0.507& 0.720& 2.012& 2.365& 749& 4.618& 4.5& 723& 870& 0.548& 0.520& 0.714& 2.033& 2.264& 782& 4.252& 4.0& 769\\ \hline
867& 0.562& 0.510& 0.716& 2.026& 2.424& 721& 4.149& 4.0& 715& 1007& 0.547& 0.518& 0.720& 2.012& 2.253& -& 3.656& 3.5& 782\\ \hline
993& 0.567& 0.516& 0.719& 2.016& 2.401& -& 3.588& 3.5& 724& & & & & & & & & &  \\ \hline
\multicolumn{10}{||c||}{$d_{pre}$(avg) = 742 nm, $d_{accu}$(avg) = 723 nm, d(exp) = 750 nm} &
\multicolumn{10}{c||}{$d_{pre}$(avg) = 817 nm, $d_{accu}$(avg) = 784 nm, d(exp) = 790 nm}\\ \hline\hline
\end{tabular}
\end{table*}
\begin{figure}
  \centering
  \includegraphics[height=12.2cm, width=9.5cm]{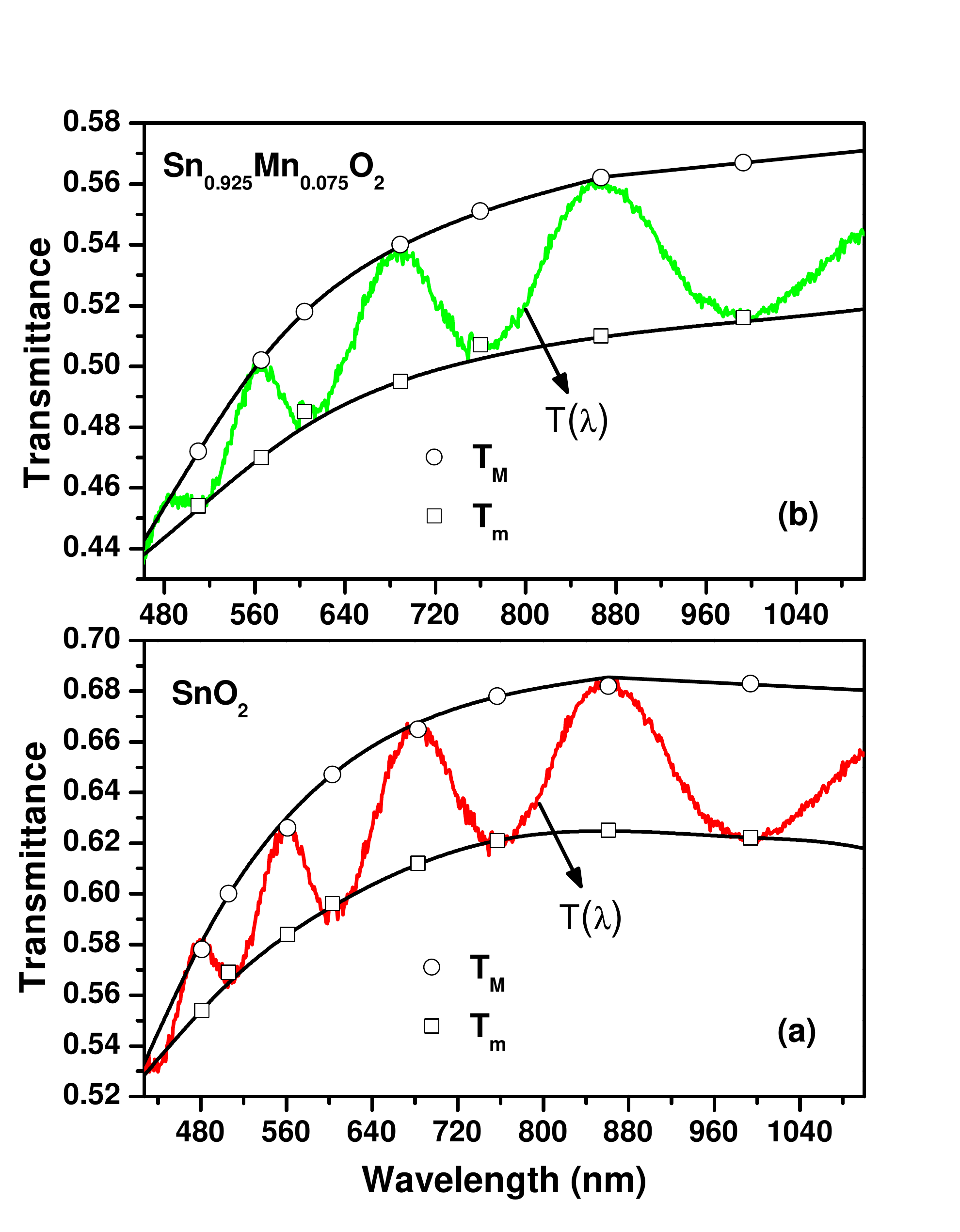}\\
  \caption{Typical transmittance spectra for two thin films $SnO_{2}$ and $Sn_{0.925}Mn_{0.075}O_{2}$. Curves $T_{M}$ and $T_{m}$, according to the text.}\label{16}
\end{figure}

The thickness of $Sn_{1-x}Mn_{x}O_{2}$ films can be calculated from transmittance data using the method proposed by Swanepoel [104]. The applicability of this method is limited to thin films deposited on transparent substrates much thicker than the thin film, conditions met in this study. The application of this method entails, as a first step, the calculation of the maximum $T_{M}(\lambda)$ and minimum $T_{m}(\lambda)$ transmittance envelope curves by parabolic interpolation to the experimentally determined positions of peaks and valleys.

From $T_{M}(\lambda)$ and $T_{m}(\lambda)$, the refractive index of the film $n(\lambda)$ in the spectral domain of the medium and strong transmission can then be calculated by the expression [104]:\\\\
$n (\lambda) = [(\frac{2s(T_{M}(\lambda) - T_{m}(\lambda))}{T_{M}(\lambda) T_{m}(\lambda)} + \frac{s^{2}+1}{2})$
\begin{equation}\label{32}
+ \sqrt{(\frac{2s(T_{M}(\lambda) - T_{m}(\lambda))}{T_{M}(\lambda) T_{m}(\lambda)} + \frac{s^{2}+1}{2})^{2} - s^{2}}]^{1/2}
\end{equation}
with s being the refractive index of the substrate. In general, s is evaluated from the transmittance spectrum of the bare substrate by the expression:
\begin{equation}\label{32}
s = \frac{1}{T_{s}} + \sqrt{\frac{1}{T_{s}^{2}} - 1}
\end{equation}
where $T_{s}$ is the substrate transmittance, which is almost a constant in the transparent zone ($\lambda$ $>$ 400 nm). The values of the refractive index $n(\lambda)$ in the $\lambda$ = 350-1000 nm range, as calculated from Eq. (17) are shown in Table IV.

If n($\lambda_{1}$) and n($\lambda_{2}$) are the refractive indices calculated from two consecutive maxima or minima corresponds to two wavelengths of $\lambda_{1}$ and $\lambda_{2}$, then the thickness of the film d can be obtained from [104]:
\begin{equation}\label{32}
d = \frac{\lambda_{1} \times \lambda_{2}}{2[\lambda_{1} n(\lambda_{2}) - \lambda_{2} n(\lambda_{1})]}
\end{equation}
The values of thickness d of the studied films determined by this equation are listed as $d_{pre}$ in Table IV.

Practically, there will be errors in the determination of extreme positions and the corresponding values of $T_{M}(\lambda)$ and $T_{m}(\lambda)$. Therefore, the preliminary values of the film thickness obtained from Eq. (19), to be denoted respectively by $d_{pre}$, are inaccurate. The more accurate film thickness can be obtained by further performing the following steps. Firstly, take the average value of $d_{pre}$ obtained from each two adjacent maxima or minima. Secondly, use the basic equation for the interference fringes 2nd = m$\lambda$ to determine the estimated order number ($m_{est}$) for each maxima or minima from the average value of $d_{pre}$ along with $n(\lambda)$ and round off each resulting $m_{est}$ to the closest half integer for minima or integer for maxima. These round values will be considered as the exact order number $m_{exact}$ corresponding to each extreme. Finally, use $m_{exact}$ and $n(\lambda)$ again to calculate the accurate thickness $d_{accurate}$ for each extreme. The average value of $d_{accurate}$ will be taken as the final thickness of the film. The values of $d_{accurate}$ found in this way have a smaller dispersion ($\sigma_{pre} > \sigma_{accurate}$).
\subsection{Magnetic properties}
Measurements of the sample magnetization as a function of temperature [M(T)] and magnetic field [M(H)] have been carried out over a temperature range of 5-300 K and field range of 0 to $\pm$ 2 T using a SQUID magnetometer. Figs. 12 and 13 show the magnetization versus applied magnetic field (M-H) curves measured at 5 and 300 K for the $Sn_{0.975}Mn_{0.025}O_{2}$, $Sn_{0.950}Mn_{0.050}O_{2}$, $Sn_{0.925}Mn_{0.075}O_{2}$ and $Sn_{0.875}Mn_{0.125}O_{2}$ films with n = $2.251 \times 10^{20}$, $2.135 \times 10^{20}$, $1.353 \times 10^{20}$ and $1.225 \times 10^{20}$ electrons $cm^{-3}$ respectively. The magnetic field was applied parallel to the film plane. The inset of Figs. 12 and 13 shows a zoom of the region of low magnetic fields that evidences the presence of a hysteresis. The saturation magnetization $(M_{S})$ is estimated to be $28.076\times10^{-2}$, $25.374\times10^{-2}$, $14.625\times10^{-2}$ and $11.388\times10^{-2}$ emu/g at 5 K and $22.694\times10^{-2}$, $19.687\times10^{-2}$, $8.878\times10^{-2}$ and $5.743\times10^{-2}$ emu/g at 300 K for the $Sn_{0.975}Mn_{0.025}O_{2}$, $Sn_{0.950}Mn_{0.050}O_{2}$, $Sn_{0.925}Mn_{0.075}O_{2}$ and $Sn_{0.875}Mn_{0.125}O_{2}$ films by the M-H curves, respectively. The thin films with higher carrier concentration ($Sn_{0.975}Mn_{0.025}O_{2}$) show ferromagnetic characteristics with higher saturation magnetization.
\begin{figure}
  \centering
  \includegraphics[height=7.3cm, width=9.5cm]{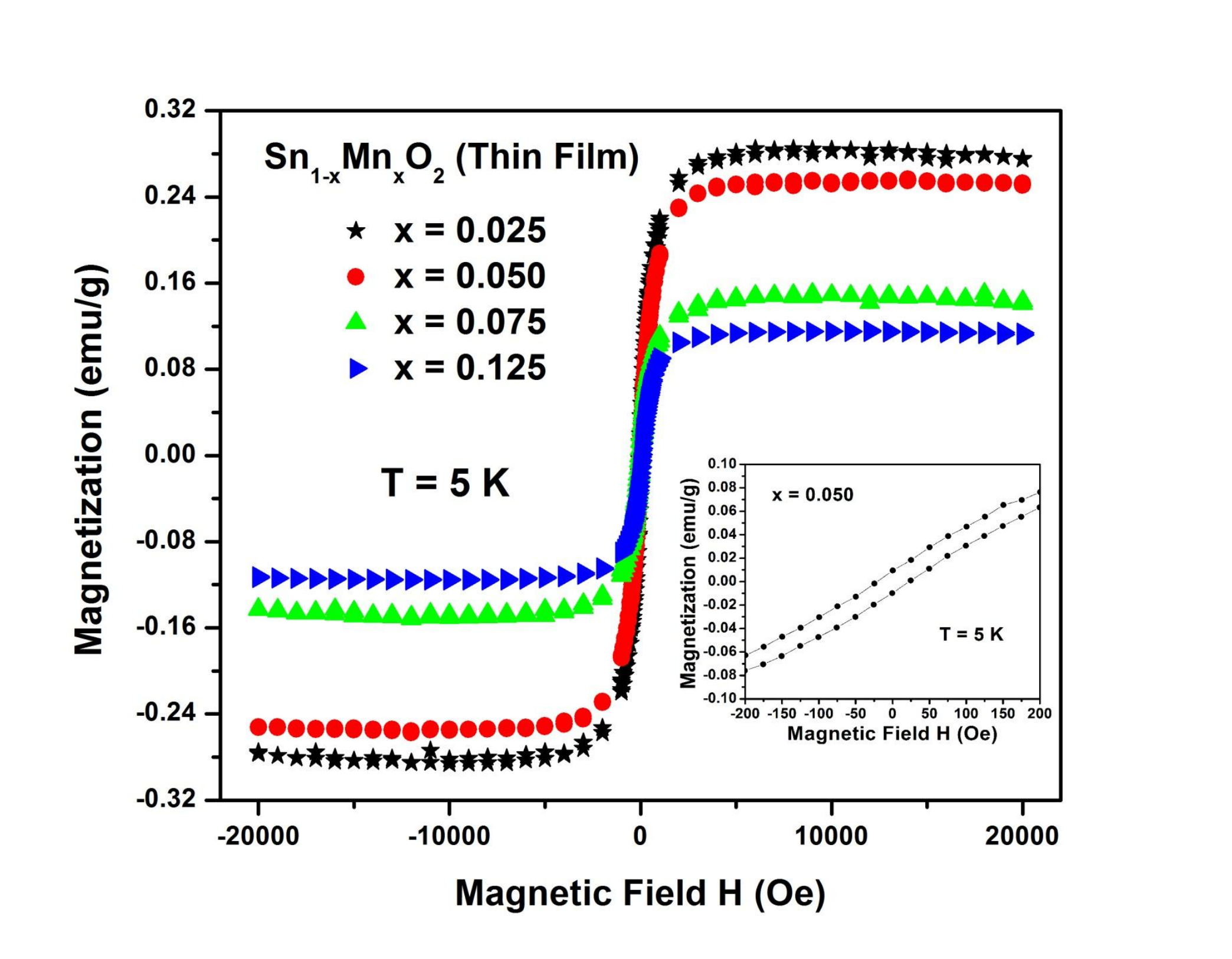}\\
  \caption{Field-dependent magnetization of the $Sn_{1-x}Mn_{x}O_{2}$ (x = 0.025, 0.050, 0.075 and 0.125) thin films measured at 5K. The inset shows the low-field part in an enlarged scale that evidences the presence of a hysteresis in $Sn_{0.950}Mn_{0.050}O_{2}$ sample.}\label{17}
\end{figure}
\begin{figure}
  \centering
  \includegraphics[height=7.3cm, width=9.5cm]{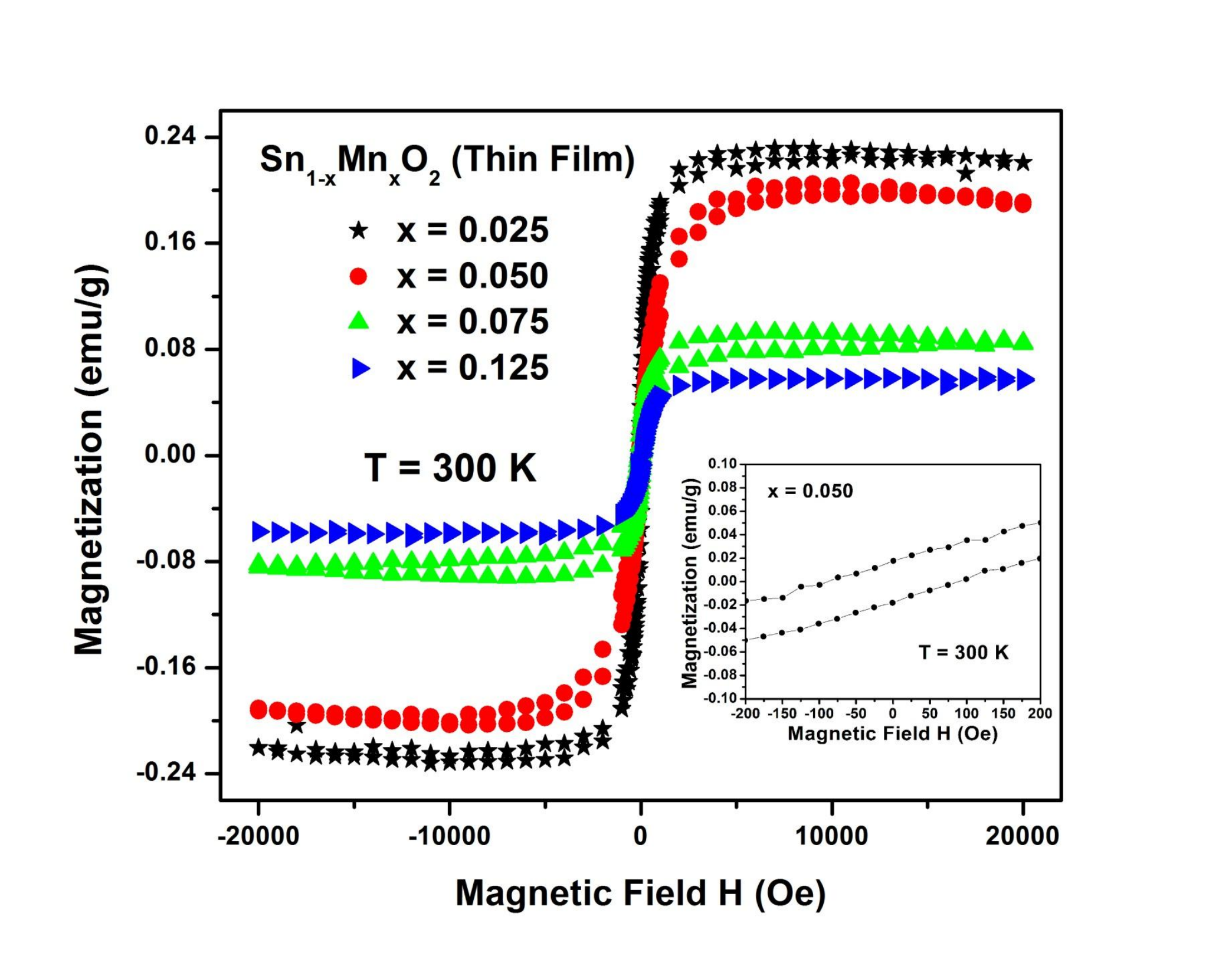}\\
  \caption{Field-dependent magnetization of the $Sn_{1-x}Mn_{x}O_{2}$ (x = 0.025, 0.050, 0.075 and 0.125) thin films measured at 300K. The inset shows the low-field part in an enlarged scale that evidences the presence of a hysteresis in $Sn_{0.950}Mn_{0.050}O_{2}$ sample.}\label{18}
\end{figure}
Since $SnO_{2}$ is an intrinsic n-type semiconductor and Mn acts as $Mn^{3+}$ in $Sn_{1-x}Mn_{x}O_{2}$ with x = 0.075, 0.100 and 0.125, according to our analysis on the lattice parameters and Hall measurement data, holes would be introduced by the $Mn^{3+}$ replacing $Sn^{4+}$, which would annihilate part of the intrinsic n-type carriers and decrease the density of carriers. Differently, for the thin films with x = 0.025 and 0.050, Mn acts as $Mn^{4+}$, thus no n-type carriers were annihilated by the isovalent ion substitution in principle. Therefore, as the doped Mn content increases (from 5\% to 7.5\%), the carrier density and accordingly the carrier mediated ferromagnetic Ruderman-Kittel-Kasuya-Yosida (RKKY) interaction decrease. According to RKKY theory, the observed magnetic properties are due to the exchange interaction between local spin-polarized electrons (such as the localized inner d-shell electrons of $Mn^{3+}$ and $Mn^{4+}$ ions) and conduction electrons. The conduction electrons are regarded as a media to interact among the $Mn^{3+}$/$Mn^{4+}$ ions.
\begin{figure}
  \centering
  \includegraphics[height=7.3cm, width=9.5cm]{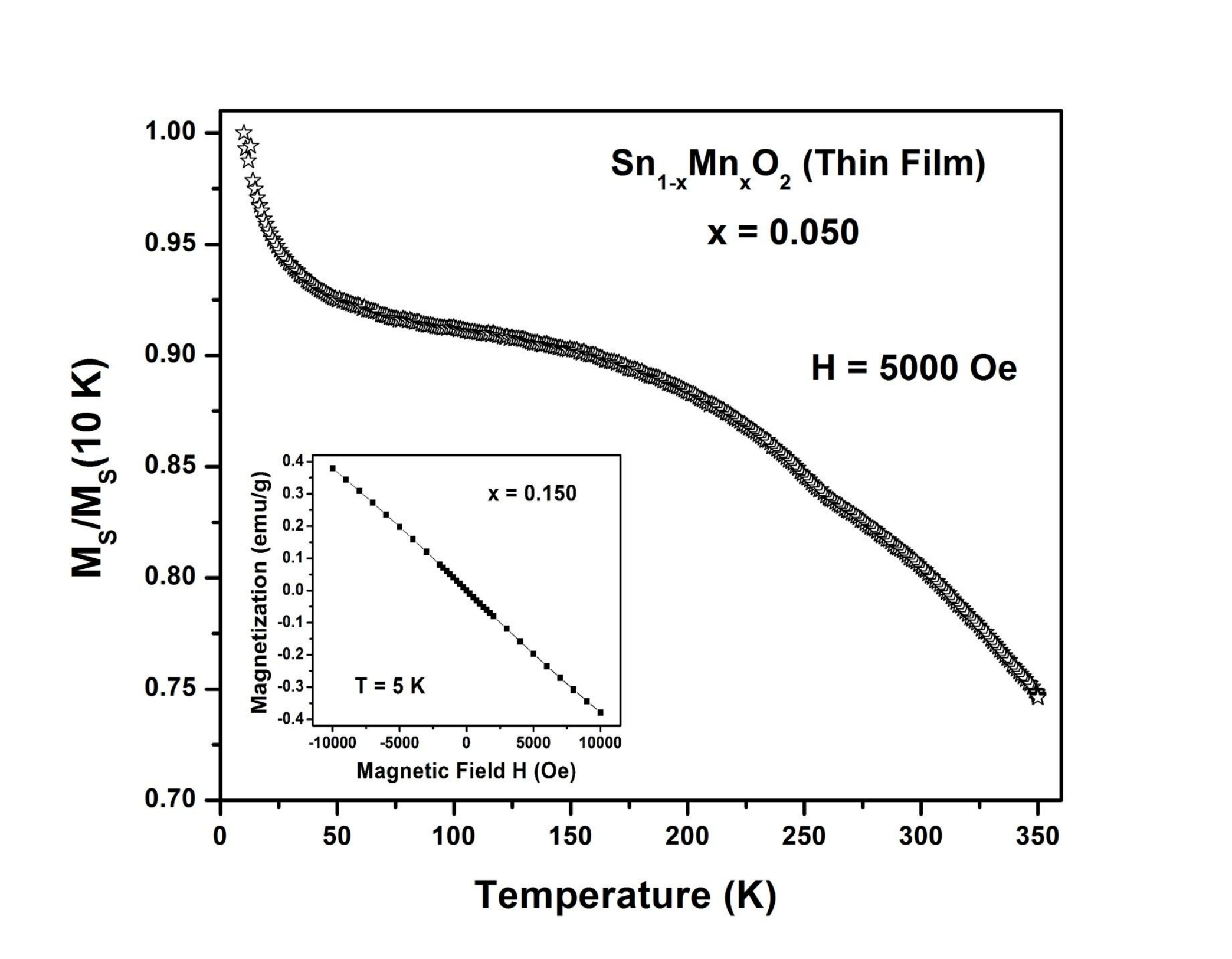}\\
  \caption{Normalized $M_{S}(T)$ plot with H = 5000 Oe for $Sn_{0.950}Mn_{0.050}O_{2}$ thin film. The inset shows the 5K field-dependent magnetization of $Sn_{0.850}Mn_{0.150}O_{2}$ thin film.}\label{19}
\end{figure}
A magnetic Mn ion induces a spin polarization in the conduction electrons in its neighborhood. This spin polarization in the itinerant electrons is felt by the moments of other magnetic Mn ions within the range leading to an indirect coupling. The saturation magnetization thus seems to have a close connection with the carrier concentration and the oxidation state of the dopant. However, we do not rule out the possibility of any other mechanism as a cause of the observed RTFM in the $Sn_{1-x}Mn_{x}O_{2}$ system. Another plausible mechanism, which has been proposed by Park et. al. [105] to explain the ferromagnetism in DMS is an $H_{o}^{+}$ defect mediated ferromagnetic spin-spin interaction. The $H_{o}^{+}$-mediated spin-spin interaction between the magnetic Mn ions is a short-range nearest-neighbor interaction that occurs through direct bonding of hydrogen to two neighboring magnetic Mn ions. Hydrogen in the Mn-$H_{o}^{+}$-Mn complex is much more stable when the neighboring $Mn^{3+}$/$Mn^{4+}$ spins are parallel rather than antiparallel, thus, giving room-temperature ferromagnetism in $Sn_{1-x}Mn_{x}O_{2}$ system.
\begin{figure}
  \centering
  \includegraphics[height=7.3cm, width=9.5cm]{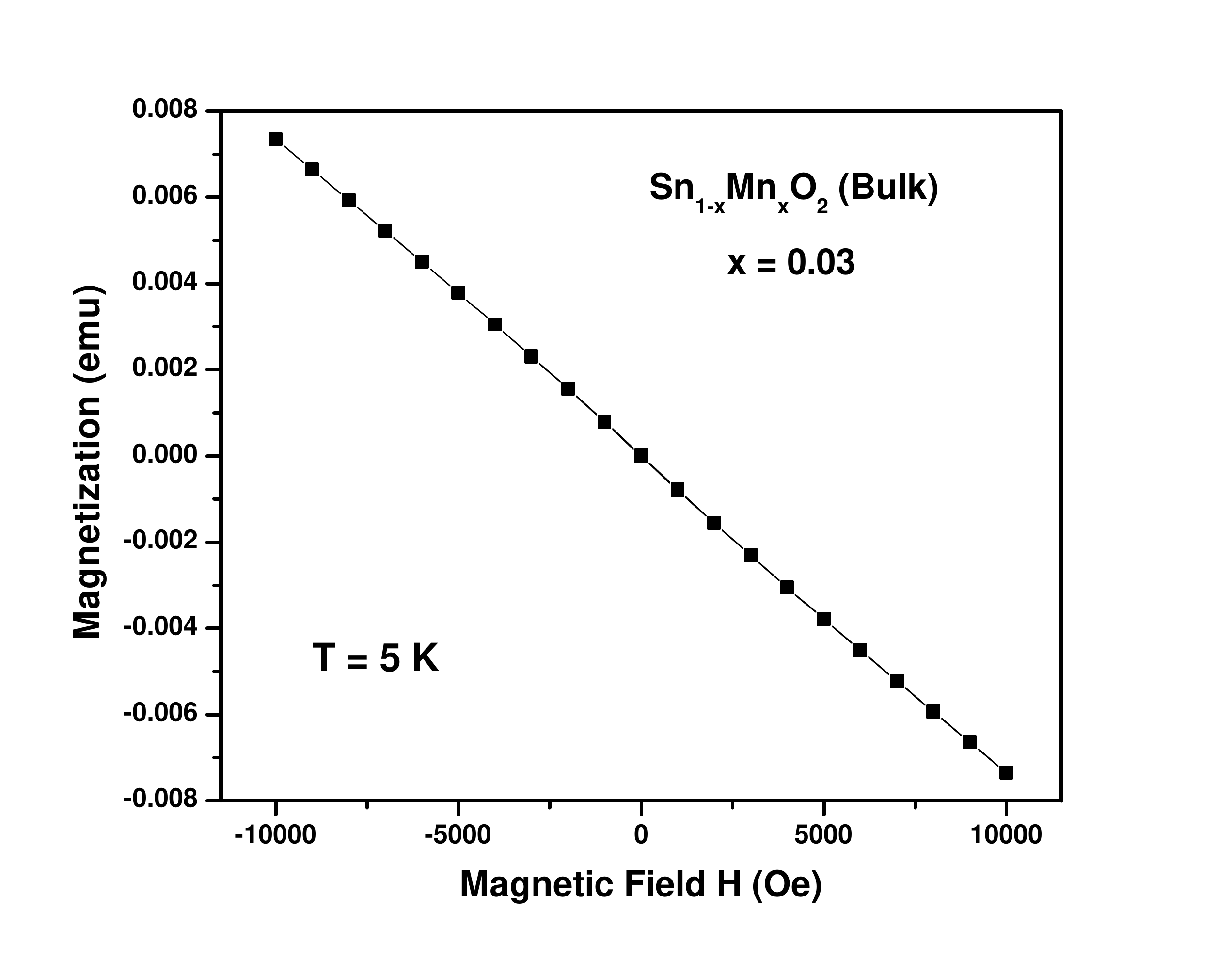}\\
  \caption{Field-dependent magnetization of the $Sn_{0.97}Mn_{0.03}O_{2}$ bulk sample measured at 5K. }\label{18}
\end{figure}

For the $Sn_{0.975}Mn_{0.025}O_{2}$ film, the M-H curve at 5 K differed by less than 19\% from those measured at room temperature (300 K), from which we guess that the Curie temperature $(T_{C})$ of $Sn_{0.975}Mn_{0.025}O_{2}$ sample is well above room temperature. But for the $Sn_{0.950}Mn_{0.050}O_{2}$, $Sn_{0.925}Mn_{0.075}O_{2}$ and $Sn_{0.875}Mn_{0.125}O_{2}$ films, the M-H curves at 5 K differed by 22\%, 39\% and 50\% from those measured at room temperature, respectively. This clearly indicates that the Curie temperatures of films are decreasing on increasing Mn doping. We also recorded the M vs T curves of these samples in a field of 0.5 T. Fig. 14 displays the M vs T plot for $Sn_{0.950}Mn_{0.050}O_{2}$ thin film. The absence of any sharp drop in the M vs T curve of $Sn_{0.950}Mn_{0.050}O_{2}$ thin film suggests that the film is ferromagnetic with a Curie temperature exceeding 350 K. For Mn concentration exceeding 12.5 at.\%, the films exhibited diamagnetic behavior. The absence of ferromagnetism for x = 0.150 is due to the possibility that now less Mn ions are incorporated in the $SnO_{2}$ lattice, as evidenced from the Hall and structural measurements, thus causing the ferromagnetism to dissappear. Magnetic measurements carried out on the pure $SnO_{2}$ films showed the expected diamagnetism with a negative magnetic susceptibility. The diamagnetic background of the pure $SnO_{2}$ film and substrate has been subtracted from all of the magnetization data shown here. Here, it is worthwhile mentioning that our magnetic measurements carried out on the bulk $Sn_{1-x}Mn_{x}O_{2}$ showed the expected diamagnetism with a negative magnetic susceptibility (see Fig. 15). The defects and free-carrier density are the important factors for ferromagnetism in Mn doped $SnO_{2}$. In the present case, ferromagnetism in bulk samples has not been observed because of the large formation energy of defects in bulk.

The presence of room temperature ferromagnetism in $Sn_{1-x}Mn_{x}O_{2}$ (x = 0.000 to 0.125) films cannot be due to the existence of secondary phases. Because the metallic manganese and almost all of the possible manganese-based binary and ternary oxide phases (MnO, $MnO_{2}$ and $Mn_{2}O_{3}$) are antiferromagnetic with Neel temperature that is much less than 300 K. However, $SnMn_{2}O_{4}$ and $Mn_{3}O_{4}$ phases are exceptions; they are ferromagnetic with Curie temperatures of 46 K and 53 K, respectively [106-108]. In the present work, the electron and x-ray diffraction analyses have not revealed any manganese oxide phases, although x-ray diffraction technique is not sensitive enough to detect secondary phases, if present at a very minute level. Even if these ferromagnetic $SnMn_{2}O_{4}$ and $Mn_{3}O_{4}$ phases are present, these cannot responsible for the ferromagnetic behavior appeared at room temperature in $Sn_{1-x}Mn_{x}O_{2}$ thin films.
\section{Conclusions}
Highly conducting Mn-doped tin oxide thin films were successfully deposited by spray pyrolysis technique on glass substrates at 450$^{o}$C. The analysis of X-ray diffraction patterns reveals that all Mn-doped tin oxide thin films are pure crystalline with tetragonal rutile phase of tin oxide which belongs to the space group P$4_{2}$/mnm (number 136). The Williamson-Hall (W-H) method has been used to evaluate the crystallite size and the microstrain of all the thin films. The average crystallite size of $Sn_{1-x}Mn_{x}O_{2}$ (x = 0.000, 0.025, 0.050, 0.075, 0.100 and 0.125) nanoparticles estimated from W-H analysis and TEM analysis is highly inter-correlated. Typical TEM micrographs of $Sn_{1-x}Mn_{x}O_{2}$ (x = 0.000, 0.025, 0.050, 0.075, 0.100 and 0.125) thin films show well isolated spherical shaped crystallites. Electron diffraction patterns taken from several crystallites confirm the $SnO_{2}$ structure in $Sn_{1-x}Mn_{x}O_{2}$ (x = 0.000, 0.025, 0.050, 0.075, 0.100 and 0.125) thin films and no evidence for secondary phases are observed. Electrical measurement shows that Mn-doped tin oxide thin films are in conducting state. The results of electrical measurements suggest that $H_{O}^{+}$ defects in Mn-doped $SnO_{2}$ thin films are responsible for the conductivity. Through electrical investigation it has also been found that the main scattering mechanism reducing the intra-grain mobility of the electrons in as-deposited $SnO_{2}$ thin films is the ionized impurity scattering. Ionized impurity scattering with singly ionized $H_{O}^{+}$ donor best describes the mobility of as-deposited $SnO_{2}$ thin films. The optical band gap ($E_{g}$) of the Mn-doped tin oxide thin films has been determined from the spectral dependence of the absorption coefficient $\alpha$ by the application of conventional extrapolation method (Tauc plot). With increasing Mn concentrations, the optical bandgap of the compounds shows a redshift compared to the host oxide $SnO_{2}$. The decrease in bandgap for increasing Mn content is attributed to the strong exchange interactions between sp carriers of host $SnO_{2}$ and localized d electrons of Mn dopant. The average transmittance of the Mn-doped $SnO_{2}$ films ranges from 50\% to 65\% (substrate transmittance $\sim$ 71\%). The magnetization as a function of magnetic field showed hysteretic behavior at room temperature. According to the temperature dependence of the magnetization, the Curie temperature is higher than 350 K. Ferromagnetic thin films of Mn-doped $SnO_{2}$ exhibited low electrical resistivity  and high optical transmittance in the visible region. No evidence of any impurity phases are detected in $Sn_{1-x}Mn_{x}O_{2}$ (x = 0.000, 0.025, 0.050, 0.075, 0.100 and 0.125) films suggesting that the emerging ferromagnetism in this system is most likely related to the properties of host $SnO_{2}$ system.
\begin{acknowledgments}
One of the authors (Sushant Gupta) is thankful to Inter University Accelerator Centre (IUAC), New Delhi  for awarding Research Fellowship through the UFUP
Project scheme (Project No. UFR-49311). The authors gratefully acknowledge to D. Kanjilal, F. Singh, A. Tripathi, K. Asokan, P. K. Kulriya, IUAC, New Delhi and A. Banerjee, M. Gupta, M. Gangrade, UGC-DAE Consortium for Scientific Research, Indore for providing the characterization facilities.
\end{acknowledgments}


\begin{thebibliography}{99}
\bibitem{ref1}
K. L. Chopra, S. Mayor, D. K. Pandya, Transparent conductors - a status review, Thin Solid Films 102 (1983) 1-46.
\bibitem{ref2}
H. L. Hartnagel, A. L. Dewar, A. K. Jain, C. Jagadish, Semiconducting Transparent Thin Films, IOP Publishing, Bristol, 1995.
\bibitem{ref3}
J. L. Vossen, Transparent conducting electrodes, in: G. Hass, M. H. Francombe, R. W. Hoffman (Eds.), Physics of Thin Films, Vol. 9, Academic Press, New York, 1976.
\bibitem{ref4}
B. G. Lewis, D. C. Paine, Applications and processing of transparent conducting oxides, MRS Bull. 25 (2000) 22-27.
\bibitem{ref5}
S. D. Bader, S. S. P. Parkin, Spintronics, Annual Review of Condensed Matter Physics 1 (2010) 71-88.
\bibitem{ref6}
T. Dietl, Origin of ferromagnetic response in diluted magnetic semiconductors and oxides, J. Phys.: Condens. Matter 19 (2007) 165204:1-15.
\bibitem{ref7}
S. J. Pearton, C. R. Abernathy, M. E. Overberg, G. T. Thaler, D. P. Norton, N. Theodoropoulou, A. F. Hebard, Y. D. Park, F. Ren, J. Kim, L. A. Boatner, Wide band gap ferromagnetic semiconductors and oxides, J. Appl. Phys.  93 (2003) 1-13.
\bibitem{ref8}
T. Fukumura, H. Toyosaki, Y. Yamada, Magnetic oxide semiconductors, Semicond. Sci. Technol. 20 (2005) S103-S111.
\bibitem{ref9}
Ranber Singh, Unexpected magnetism in nanomaterials, J. Magn. Magn. Mater. 346 (2013) 58-73.
\bibitem{ref10}
S. A. Wolf, D. D. Awschalom, R. A. Buhrman, J. M. Daughton, S. von Moln\'{a}r, M. L. Roukes, A. Y. Chtchelkanova, D. M. Treger, Spintronics: a spin-based electronics vision for the future,  Science 294 (2001) 1488-1495.
\bibitem{ref11}
W. Prellier, A. Fouchet, B. Mercey, Oxide-diluted magnetic semiconductors: a review of the experimental status, J. Phys. Condens. Matter 15 (2003) R1583-R1601.
\bibitem{ref12}
J. M. D. Coey, M. Venkatesan, C. B. Fitzgerald, Donor impurity band exchange in dilute ferromagnetic oxides, Nature Materials 4 (2005) 173-179.
\bibitem{ref13}
R. C. Budhani, P. Pant, R. K. Rakshit, K. Senapati, S. Mandal, N. K. Pandey, J. Kumar,  Magnetotransport in epitaxial films of the degenerate semiconductor $Zn_{1-x}Co_{x}O$, J. Phys. Condens. Matter 17 (2005) 75.
\bibitem{ref14}
A. F. Orlov, L. A. Balagurov, A. S. Konstantinova, N. S. Perov, D. G. Yarkin, Giant magnetic moments in dilute magnetic semiconductors, J. Magn. Magn. Mater. 320 (2008) 895-897.
\bibitem{ref15}
S. B. Ogale, R. J. Choudhary, J. P. Buban, S. E. Lofland, S. R. Shinde, S. N. Kale, V. N. Kulkarni, J. Higgins, C. Lanci, J. R. Simpson, N. D. Browning, S. Das Sarma, H. D. Drew, R. L. Greene, T. Venkatesan, High temperature ferromagnetism with a giant magnetic moment in transparent Co-doped $SnO_{2-\delta}$,  Phys. Rev. Lett.  91 (2003) 077205:1-4.
\bibitem{ref16}
H. H. Nguyen, J. Sakai, N. T. Huong, N. Poirot, A. Ruyter, Role of defects in tuning ferromagnetism in diluted magnetic oxide thin films, Phys. Rev. B 72 (2005) 045336:1-5.
\bibitem{ref17}
N. H. Hong, A. Ruyter, W. Prellier, J. Sakai, N. T. Huong, Magnetism in Ni-doped $SnO_{2}$ thin films, J. Phys.: Condens. Matter 17 (2005) 6533-6538.
\bibitem{ref18}
N. H. Hong, J. Sakai, Ferromagnetic V-Doped $SnO_{2}$ thin films, Physica B 358 (2005) 265-268.
\bibitem{ref19}
N. H. Hong, J. Sakai, W. Prellier, A. Hassini, Transparent Cr-doped $SnO_{2}$ thin films: ferromagnetism beyond room temperature with a giant magnetic moment, J. Phys.: Condens. Matter  17 (2005) 1697-1702.
\bibitem{ref20}
J. M. D. Coey, A. P. Douvalis, C. B. Fitzgerald, M. Venkatesan,  Ferromagnetism in Fe-doped $SnO_{2}$ thin films,  Appl. Phys. Lett. 84 (2004) 1332-1334.
\bibitem{ref21}
A. Punnoose, J. Hays, A. Thurber, M. H. Engelhard, R. K. Kukkadapu, C. Wang, V. Shutthanandan, S. Thevuthasan, Development of high-temperature ferromagnetism in $SnO_{2}$ and paramagnetism in SnO by Fe doping, Phys. Rev. B 72 (2005) 054402:1-14.
\bibitem{ref22}
K. Gopinadhan, S. C. Kashyap, D. K. Pandya, S. Chaudhary, High temperature ferromagnetism in Mn-doped $SnO_{2}$ nanocrystalline thin films, J. Appl. Phys. 102 (2007) 113513:1-8.
\bibitem{ref23}
C. B. Fitzgerald, M. Venkatesan, A. P. Douvalis, S. Huber, J. M. D. Coey, T. Bakas, $SnO_{2}$ doped with Mn, Fe or Co: Room temperature dilute magnetic semiconductors, J. Appl. Phys. 95 (2004) 7390-7392.
\bibitem{ref24}
L. B. Duan, G. H. Rao, J. Yu, Y. C. Wang, G. Y. Liu, J. K. Liang, Structural and magnetic properties of chemically synthesized $Sn_{1-x}Mn_{x}O_{2}$ nanocrystalline powders, J. Appl. Phys. 101 (2007) 063917:1-6.
\bibitem{ref25}
H. Kimura, T. Fukumura, M. Kawasaki, K. Inaba, T. Hasegawa, H. Koinuma, Rutile-type oxide-diluted magnetic semiconductor: Mn-doped $SnO_{2}$, Appl. Phys. Lett. 80 (2002) 94-96.
\bibitem{ref26}
Sushant Gupta, F. Singh, N. P. Lalla, B. Das, Swift heavy ion irradiation induced modifications in structural, microstructural, electrical and magnetic properties of Mn doped $SnO_{2}$ thin films, Nucl. Instrum. Methods Phys. Res. B 400 (2017) 37-57.
\bibitem{ref27}
A. M. Abdel Hakeem, Structure and magnetic properties of $Sn_{1-x}Mn_{x}O_{2}$, J. Magn. Magn. Mater. 324 (2012) 95-99.
\bibitem{ref28}
A. Punnoose, J. Hays, V. Gopal, V. Shutthanandan, Room-temperature ferromagnetism in chemically synthesized $Sn_{1-x}Co_{x}O_{2}$ powders, Appl. Phys. Lett. 85 (2004) 1559-1561.
\bibitem{ref29}
K. Nomura, C. A. Barrero, J. Sakuma, M. Takeda, Room-temperature ferromagnetism of sol-gel synthesized $Sn_{1-x}Fe_{x}^{57}O_{2-\delta}$ powders, Phys. Rev. B  75 (2007) 184411:1-13.
\bibitem{ref30}
J. Hays, A. Punnoose, R. Baldner, M. H. Engelhard, J. Peloquin, K. M. Reddy, Relationship between the structural and magnetic properties of Co-doped $SnO_{2}$ nanoparticles, Phys. Rev. B 72 (2005) 075203:1-7.
\bibitem{ref31}
S.A. Ahmed, S.H. Mohamed, Room temperature ferromagnetism behavior of $Sn_{1-x}Mn_{x}O_{2}$ powders, J. Magn. Magn. Mater. 324 (2012) 812-817.
\bibitem{ref32}
C. B. Fitzgerald, M. Venkatesan, L. S. Dorneles, R. Gunning, P. Stamenov, J. M. D. Coey, P. A. Stampe, R. J. Kennedy, E. C. Moreira, U. S. Sias, Magnetism in dilute magnetic oxide thin films based on $SnO_{2}$, Phys. Rev. B 74 (2006) 115307-115317.
\bibitem{ref33}
W. Wang, Z. Wang, Y. Hong, J. Tang, M. Yu, The structure and magnetic properties of Cr/Fe-doped $SnO_{2}$ thin films,  J. Appl. Phys. 99 (2006) 08M115:1-3.
\bibitem{ref34}
V. Bilovol, A. F. Cabrera, C. E. Rodriguez Torres, A. M. Mudarra Navarro, Study of magnetic state of $Sn_{0.9}Fe_{0.1}O_{2}$ powders at low temperature, J. Magn. Magn. Mater. 344 (2013) 188-192.
\bibitem{ref35}
K. Gopinadhan, D. K. Pandya, S. C. Kashyap, S. Chaudhary, Cobalt-substituted $SnO_{2}$ thin films: a transparent ferromagnetic semiconductor, J. Appl. Phys. 99 (2006) 126106:1-3.
\bibitem{ref36}
V. G. Kravets, L. V. Poperenko, Magnetic ordering effects in the Raman spectra of $Sn_{1-x}Co_{x}O_{2}$,  J. Appl. Phys. 103 (2008) 083904:1-6.
\bibitem{ref37}
R. Adhikari, A. K. Das, D. Karmakar, T. V. Chandrasekhar Rao, J. Ghatak, Structure and magnetism of Fe-doped $SnO_{2}$ nanoparticles, Phys. Rev. B 78 (2008) 024404:1-9.
\bibitem{ref38}
P. I. Archer, D. R. Gamelin, Controlled grain-boundary defect formation and its role in the high-$T_{c}$ ferromagnetism of $Ni^{2+}$: $SnO_{2}$, J. Appl. Phys. 99 (2006) 08M107:1-3.
\bibitem{ref39}
S. K. Misra, S. I. Andronenko, K. M. Reddy, J. Hays, A. Punnoose, Magnetic resonance studies of $Co^{2+}$ ions in nanoparticles of $SnO_{2}$ processed at different temperatures,   J. Appl. Phys. 99 (2006), 08M106:1-3.
\bibitem{ref40}
X. F. Liu, R. H. Yu, Mediation of room temperature ferromagnetism in Co-doped $SnO_{2}$ nanocrystalline films by structural defects , J. Appl. Phys. 102 (2007) 083917:1-5.
\bibitem{ref41}
W. Zhou, L. Liu, P. Wu, Nonmagnetic impurities induced magnetism in $SnO_{2}$, J. Magn. Magn. Mater. 321 (2009) 3356-3359.
\bibitem{ref42}
J. F. Liu, M. F. Lu, P. Chai, L. Fu, Z. L. Wang, X. Q. Cao, J. Meng, The magnetic and structural properties of hydrothermal-synthesized single-crystal $Sn_{1−x}Fe_{x}O_{2}$ nanograins, J. Magn. Magn. Mater. 317 (2007) 1-7.
\bibitem{ref43}
F. H. Aragón, J. A. H. Coaquira, L. C. C. M. Nagamine, R. Cohen, S. W. da Silva, P. C. Morais, Thermal-annealing effects on the structural and magnetic properties of 10\% Fe-doped $SnO_{2}$ nanoparticles synthetized by a polymer precursor method, J. Magn. Magn. Mater. 375 (2015) 74-79.
\bibitem{ref44}
S. Ghosh, M. Mandal, K. Mandal, Effects of Fe doping and Fe-N codoping on magnetic properties of $SnO_{2}$ prepared by chemical co-precipitation, J. Magn. Magn. Mater. 323 (2011) 1083-1087.
\bibitem{ref45}
S. Bhuvana, H. B. Ramalingam, K. Vadivel, E. Ranjith Kumar, Ahmad I. Ayesh, Effect of Zn and Ni substitution on structural, morphological and magnetic properties of tin oxide nanoparticles, J. Magn. Magn. Mater.  419 (2016) 429-434.
\bibitem{ref46}
C. E. Rodriguez Torres, L. Errico, F. Golmar, A. M. Mudarra Navarro, A. F. Cabrera, S. Duhalde, F. H. Sánchez, M. Weissmann, The role of the dopant in the magnetism of Fe-doped $SnO_{2}$ films, J. Magn. Magn. Mater. 316 (2007) e219-e222.
\bibitem{ref47}
A. Ali, A. K. Sarfraz, K. Ali, A. Mumtaz, Structural, optical, Induced ferromagnetism and anti-ferromagnetism in $SnO_{2}$ nanoparticles by varying cobalt concentration, J. Magn. Magn. Mater. 391 (2015) 161-165.
\bibitem{ref48}
W. Zhou, L. Liu, P. Wu, Nonmagnetic impurities induced magnetism in $SnO_{2}$, J. Magn. Magn. Mater. 321 (2009) 3356-3359.
\bibitem{ref49}
H. Wang, Y. Yan, Y. Sh. Mohammed, X. Du, K. Li, H. Jin, First-principle study of magnetism in Co-doped $SnO_{2}$, J. Magn. Magn. Mater. 321 (2009) 337-342.
\bibitem{ref50}
J. Li, G. Bai, Y. Jiang, Y. Du, C. Wu, M. Yan, Origin of room temperature ferromagnetism in $SnO_{2}$ films, J. Magn. Magn. Mater. 426 (2017) 545-549.
\bibitem{ref51}
Hyun-Suk Kim, L. Bi, G. F. Dionne, C. A. Ross, Han-Jong Paik, Structure, magnetic and optical properties, and Hall effect of Co- and Fe-doped $SnO_{2}$ films, Phys. Rev. B  77 (2008) 214436:1-7.
\bibitem{ref52}
N. Labedeva, P. Kuivalainen, Modeling of ferromagnetic semiconductor devices for spintronics, J. Appl. Phys. 93 (2003) 9845-9864.
\bibitem{ref53}
G. A. Prinz, Magnetoelectronics, Science 282 (1998) 1660:1-3.
\bibitem{ref54}
S. A. Chambers, R. F. C. Farrow, New possibilities for ferromagnetic semiconductors, MRS Bull. 28 (2003) 729-734.
\bibitem{ref55}
G. A. Prinz, Magnetoelectronics applications, J. Magn. Magn. Mater.  200 (1999) 57-68.
\bibitem{ref56}
G. K. Williamson, W. H. Hall, X-ray line broadening from filed Al and W, Acta Metall. 1 (1953) 22-31.
\bibitem{ref57}
B. D. Cullity, Elements of X-ray Diffraction, Addison-Wesley Publishing Company Inc., California, 1956.
\bibitem{ref58}
D. Szczuko, J. Werner, S. Oswald, G. Behr, K. Wetzig, XPS investigations of surface segregation of doping elements in $SnO_{2}$, Appl. Surf. Sci. 179 (2001) 301-306.
\bibitem{ref59}
R. Edson, M. Leite, B. Ins, Bernardi, Elson Longo, A. Jos, Varela, A. Carlos, Paskocimas, Enhanced electrical property of nanostructured Sb-doped $SnO_{2}$ thin film processed by soft chemical method, Thin Solid Films 449 (2004) 67-72.
\bibitem{ref60}
Jianrong Zhang, Lian Gao, Synthesis and characterization of antimony-doped tin oxide (ATO) nanoparticles, Inorg. Chem. Commun. 7 (2004) 91-93.
\bibitem{ref61}
D. Szczuko, J. Werner, G. Behr, S. Oswald, K. Wetzig, Surface-related investigations to characterize different preparation techniques of Sb-doped $SnO_{2}$ powders, Surf. Interface Anal. 31 (2001) 484-491.
\bibitem{ref62}
Wang Jianhua, Peng Guanghuai, Guo Yuzhong, Yang Xikun, XPS investigation of segregation of Sb in $SnO_{2}$ powders, J. Wuhan Univ. Technol. Mater. Sci. Ed. 23 (2008) 95-99.
\bibitem{ref63}
R. Kirchheim, Grain coarsening inhibited by solute segregation, Acta Mater. 50 (2002) 413-419.
\bibitem{ref64}
C. S. Barrett, T. B. Massalski, Structure of Metals: Crystallographic Methods, Principles and Data, third ed., McGraw-Hill, New York, 1966, p. 205.
\bibitem{ref65}
K. H. Kim, J. S. Chun, X-ray studies of $SnO_{2}$ prepared by chemical vapour deposition, Thin Solid Films 141 (1986) 287-295.
\bibitem{ref66}
H. R. Moutinho, M. M. Al-Jassim, D. H. Levi, P. C. Dippo, L.L. Kazmerski, Effects of $CdCl_{2}$ treatment on the recrystallization and electro-optical properties of CdTe thin films, J. Vac. Sci. Technnol, 16 (1998) 1251-1257.
\bibitem{ref67}
\c{C}. Kili\c{c}, A. Zunger, Origins of coexistence of conductivity and transparency in $SnO_{2}$, Phys. Rev. Lett.  88 (2002) 095501:1-4.
\bibitem{ref68}
Sushant Gupta, Fouran Singh, Indra Sulania, B. Das, Role of carrier concentration in swift heavy ion irradiation induced surface modifications, arXiv:1612.05150 [cond-mat.mtrl-sci] (2016) 1-13.
\bibitem{ref69}
M. D. McCluskey, M. C. Tarun, S. T. Teklemichael, Hydrogen in oxide semiconductors, J. Mater. Res. 27 (2012) 2190-2198.
\bibitem{ref70}
A. K. Singh, A. Janotti, M. Scheffler, C. G. Van de Walle, Sources of electrical conductivity in $SnO_{2}$, Phys. Rev. Lett.  101 (2008) 055502:1-4.
\bibitem{ref71}
A. Janotti, J. B. Varley, J. L. Lyons, C. G. Van de Walle, Controlling the conductivity in oxide semiconductors, In: J. Wu et. al. (eds.), Functional Metal Oxide Nanostructures, Springer Series in Materials Science, 2012.
\bibitem{ref72}
A. Janotti, C. G. Van de Walle, Native point defects in ZnO, Phys. Rev. B 76 (2007) 165202-165224.
\bibitem{ref73}
A. Janotti, C. G Van de Walle, Fundamentals of zinc oxide as a semiconductor, Rep. Prog. Phys. 72 (2009) 126501:1-29.
\bibitem{ref74}
A. Janotti, C. G. Van de Walle, Oxygen vacancies in ZnO, Appl. Phys. Lett.  87 (2005) 122102:1-3.
\bibitem{ref75}
A. Janotti, C. G. Van de Walle, New insights into the role of native point defects in ZnO, J. Cryst. Growth 287 (2006) 58-65.
\bibitem{ref76}
\c{C}. Kili\c{c}, A. Zunger, n-type doping of oxides by hydrogen, Appl. Phys. Lett. 81 (2002) 73-75.
\bibitem{ref77}
K. Xiong, J. Robertson, S. J. Clark, Behavior of hydrogen in wide band gap oxides, J. Appl. Phys. 102 (2007) 083710:1-13.
\bibitem{ref78}
C. G. Van de Walle, Hydrogen as a shallow center in semiconductors and oxides, Phys. Status Solidi B 235 (2003) 89-95.
\bibitem{ref79}
A. Janotti, C. G. Van de Walle, Hydrogen multicentre bonds, Nature Materials  6 (2007) 44-47.
\bibitem{ref80}
D. G. Thomas, J. J. Lander, Hydrogen as a donor in zinc oxide, J. Chem. Phys. 25 (1956) 1136-1142.
\bibitem{ref81}
C. G. Van de Walle, Hydrogen as a cause of doping in zinc oxide, Phys. Rev. Lett. 85 (2000) 1012-1015.
\bibitem{ref82}
N. H. Nickel, Hydrogen migration in single crystal and polycrystalline zinc oxide, Phys. Rev. B  73 (2006) 195204:1-9.
\bibitem{ref83}
J. Bang, K. J. Chang, Diffusion and thermal stability of hydrogen in ZnO, Appl. Phys. Lett. 92 (2008) 132109:1-3.
\bibitem{ref84}
W. M. Hlaing Oo, S. Tabatabaei, M. D. McCluskey, J. B. Varley, A. Janotti, C. G. Van de Walle, Hydrogen donors in $SnO_{2}$ studied by infrared spectroscopy and first-principles calculations, Phys. Rev. B 82 (2010) 193201:1-4.
\bibitem{ref85}
F. Bekisli, M. Stavola, W. Beall Fowler, L. Boatner, E. Spahr, G. L\"{u}pke, Hydrogen impurities and shallow donors in $SnO_{2}$ studied by infrared spectroscopy, Phys. Rev. B  84 (2011) 035213:1-8.
\bibitem{ref86}
J. B. Varley, A. Janotti, A. K. Singh, C.G. Van de Walle, Hydrogen interactions with acceptor impurities in $SnO_{2}$: first-principles calculations, Phys. Rev. B 79 (2009) 245206.
\bibitem{ref87}
P. D. C. King, R. L. Lichti, Y. G. Celebi, J. M. Gil, R. C. Vil$\tilde{a}$o, H. V. Alberto, J. Piroto Duarte, D. J. Payne, R. G. Egdell, I. McKenzie, C. F. McConville, S. F. J. Cox, T. D. Veal, Shallow donor state of hydrogen in $In_{2}O_{3}$ and $SnO_{2}$: implications for conductivity in transparent conducting oxides,  Phys. Rev. B 80 (2009) 081201:1-4.
\bibitem{ref88}
S. F. J. Cox, E. A. Davis, S. P. Cottrell, P. J. C. King, J. S. Lord, J. M. Gil, H. V. Alberto, R. C. Vil$\tilde{a}$o, J. Piroto Duarte, N. Ayres de Campos, A. Weidinger, R. L. Lichti, S. J. C. Irvine, Experimental confirmation of the predicted shallow donor hydrogen state in zinc oxide, Phys. Rev. Lett. 86 (2001) 2601-2604.
\bibitem{ref89}
G. A. Shi, M. Saboktakin, M. Stavola, S. J. Pearton, Hidden hydrogen in as-grown ZnO, Appl. Phys. Lett. 85 (2004) 5601-5603.
\bibitem{ref90}
C. G. Van de Walle, J. Neugebauer, Universal alignment of hydrogen levels in semiconductors, insulators and solutions, Nature 423 (2003) 626-628.
\bibitem{ref91}
J. R. Bellingham, W. A. Phillips, C. J. Adkins, Electrical and optical properties of amorphous indium oxide, J. Phys.: Condens. Matter 2 (1990) 6207-6221.
\bibitem{ref92}
Sushant Gupta, B. C. Yadav, P. K. Dwivedi, B. Das, Microstructural, optical and electrical investigations of Sb-$SnO_{2}$ thin films deposited by spray pyrolysis, Materials Research Bulletin 48 (2013) 3315-3322.
\bibitem{ref93}
R. B. Hadj Tahar, T. Ban, Y. Ohya, Y. Takahashi, Tin doped indium oxide thin films: electrical properties, J. Appl. Phys. 83 (1998) 2631-2645.
\bibitem{ref94}
Sushant Gupta, The Synthesis and Characterization of Transparent Conducting Antimony Doped Tin Oxide Thin Films Deposited by Spray Pyrolysis, M.Sc. Thesis, Department of Applied Physics, Babasaheb Bhimrao Ambedkar Central University, Lucknow, India, 2012, pp. 8–18.
\bibitem{ref95}
T. Nutz, M. Haase, Wet-chemical synthesis of doped nanomaterials: optical properties of oxygen-deficient and antimony-doped colloidal $SnO_{2}$, J. Phys. Chem. B 104 (2000) 8430-8437.
\bibitem{ref96}
J. W. Orton, M. J. Powell, The Hall effect in polycrystalline and powdered semiconductors, Rep. Prog. Phys., 43 (1980) 1263-1307.
\bibitem{ref97}
V. I. Fistul and V. M. Vainshtein, Mechanism of Electron Scattering in $ln_{2}O_{3}$ Films, Sov. Phys. Solid State 8 (1967) 2769.
\bibitem{ref98}
J. G. Na, Y. R. Cho, Y. H. Kim, T. D. Lee, S. J. Park, Effects of annealing temperature on microstructure and electrical and optical properties of radio-frequency-sputtered tin-doped indium oxide films, J. Am. Ceram. Soc. 72 (1989) 698-701.
\bibitem{ref99}
V. I. Fistul, Heavily Doped Semiconductors, Plenum Press, New York, 1969, p. 86.
\bibitem{ref100}
M. Batzill, U. Diebold, The surface and materials science of tin oxide, Prog. Surf. Sci. 79 (2005) 47-154.
\bibitem{ref101}
J. Tauc, R. Grogorovici, A. Vancu, Optical properties and electronic structure of amorphous germanium, Phys. Stat. Solidi. 15 (1966) 627-637.
\bibitem{ref102}
S. Tsunekawa, T. Fukuda, A. Kasuya, Blue shift in ultraviolet absorption spectra of monodisperse $CeO_{2-x}$ nanoparticles, J. Appl. Phys. 87 (2000) 1318-1321.
\bibitem{ref103}
R. B. Bylsma, W. M. Becker, J. Kossut, U. Debska, Dependence of energy gap on x and T in $Zn_{1-x}Mn_{x}Se$: The role of exchange interaction, Phys. Rev. B 33 (1986) 8207-8215.
\bibitem{ref104}
R. J. Swanepoel, Determination of the thickness and optical constants of amorphous silicon, J. Phys. E: Sci. Instrum. 16 (1983) 1214-1222.
\bibitem{ref105}
J. K. Park, K. W. Lee, H. Kweon, C. E. Lee, Evidence of hydrogen-mediated ferromagnetic coupling in Mn-doped ZnO, Appl. Phys. Lett. 98 (2011) 102502:1-3.
\bibitem{ref106}
K. Dwight, N. Menyuk, Magnetic properties of $Mn_{3}O_{4}$ and the canted spin problem, Phys. Rev. 119 (1960) 1470-1479.
\bibitem{ref107}
D. G. Wickham, N. Menuyk, K. Dwight, Evidence for canted magnetic moments in manganous stannate $(Mn_{2}SnO_{4})$, J. Phys. Chem. Solids 20 (1961) 316-318.
\bibitem{ref108}
P. Z. Si, D. Li, J. W. Lee, C. J. Choi, Z. D. Zhang, D. Y. Geng, E. Br\"{u}ck,  Unconventional exchange bias in oxide-coated manganese nanoparticles, Appl. Phys. Lett. 87 (2005) 133122:1-3.
\end{thebibliography}
\end{document}